\documentclass[12pt, draftclsnofoot, onecolumn]{IEEEtran}
\usepackage{amsthm}
\usepackage{multirow}
\usepackage{color}
\usepackage{algorithm}
\usepackage{algpseudocode}
\newtheorem*{lemma1}{Lemma 1}
\newtheorem*{lemma2}{Lemma 2}

\usepackage[T1]{fontenc}
\usepackage{multirow}
\usepackage{longtable}

\ifCLASSINFOpdf
\else
\fi
\usepackage{amsmath}
\usepackage{amsmath,bm}
\usepackage{graphicx}
\usepackage{amsmath,amssymb,amsthm,mathrsfs,amsfonts,dsfont}
\usepackage{subfigure}
\usepackage{array}
\begin{document}
%
\title{Double-Sequence Frequency Synchronization for Wideband Millimeter-Wave Systems with Few-Bit ADCs}
%
%
%

\author{Dalin Zhu,
        Ralf Bendlin,
        Salam Akoum,
        Arunabha Ghosh,
        and~Robert~W.~Heath,~Jr.
\thanks{Dalin Zhu and Robert W. Heath, Jr. are with the Department
of Electrical and Computer Engineering, The University of Texas at Austin, Austin,
TX, 78712 USA, e-mail: \{dalin.zhu, rheath\}@utexas.edu.

Ralf Bendlin, Salam Akoum and Arunabha Ghosh are with AT\&T Labs, Austin, TX, 78759 USA, e-mail: \{ralf\_bendlin, salam\_akoum, ghosh\}@labs.att.com.

This work was supported in part by the National Science Foundation under Grant No. ECCS-1711702, CNS-1702800 and CNS-1731658 and a gift from AT\&T Labs.}}

\maketitle


\begin{abstract}
In this paper, we propose and evaluate two novel double-sequence low-resolution frequency synchronization methods in millimeter-wave (mmWave) systems. In our system model, the base station uses analog beams to send the synchronization signal with infinite-resolution digital-to-analog converters. The user equipment employs a fully digital front end to detect the synchronization signal with low-resolution analog-to-digital converters (ADCs). The key ingredient of the proposed methods is the custom designed synchronization sequence pairs, from which there exists an invertible function (a ratio metric) of the carrier frequency offset (CFO) to be estimated. We use numerical examples to show that the ratio metric is robust to the quantization distortion. Further, we analytically characterize the CFO estimation performances of our proposed designs assuming a single user. To implement our proposed methods in a multi-user scenario, we propose to optimize the double-sequence design parameters such that: (i) for each individual user, the impact of the quantization distortion on the CFO estimation accuracy is minimized, and (ii) the resulting frequency range of estimation can capture as many users' CFOs as possible. Numerical results reveal that our proposed algorithms provide a flexible means to estimate CFO in a variety of low-resolution settings.
\end{abstract}

%
\IEEEpeerreviewmaketitle

\allowdisplaybreaks

\section{Introduction}
Due to the use of large bandwidth for high-rate data communications at millimeter-wave (mmWave) frequencies \cite{rhsp}, the sampling rate of the corresponding analog-to-digital converters (ADCs) scales up, resulting in high power consumption and hardware complexity. Reducing ADC resolution is a solution to reduce implementation costs \cite{hpadc}. The use of low-precision ADCs, though, brings new design challenges to practical cellular networks. Implementing low-resolution ADCs in communications systems has been investigated in various aspects, including input signal optimization \cite{amjn,jsodum}, mutual information analysis \cite{jmrwj}-\nocite{mijn}\cite{bmic}, channel estimation \cite{tlvw}-\nocite{gzgk}\cite{jmps}, uplink multiuser detection \cite{swyl}-\nocite{jcjmrj}\cite{cmjcel}, and frame timing synchronization \cite{dztwctiming}. In practice, low-resolution quantization will also impair the frequency synchronization performance of mmWave systems \cite{awumw}. Prior work assumes that frequency synchronization is performed without quantization distortion.

There are many pilot/sequence-aided frequency synchronization methods for orthogonal frequency division multiplexing (OFDM) systems \cite{cfo0}-\nocite{cfo1}\nocite{cfo2}\nocite{cfo3}\nocite{cfo4}\cite{cfo5}. The Cox-Schmidl algorithm \cite{cfo0} and variants \cite{cfo1}-\nocite{cfo2}\nocite{cfo3}\cite{cfo4} is the classic pilot-aided carrier frequency offset (CFO) estimation approach. In the Cox-Schmidl algorithm, a first training sequence is used to estimate the CFO with an ambiguity identical to one subcarrier spacing, while a second training sequence is employed to resolve this ambiguity. In \cite{cfo5}, the symmetry of the Zadoff-Chu (ZC) sequences was exploited in the 3GPP long-term evolution (LTE) systems for frequency synchronization. Similar to the first training sequence in the Cox-Schmidl algorithm, the symmetry of the ZC sequences creates time-domain periodicity in one OFDM symbol duration to track the CFOs. That approach, however, only works for fractional CFO that is less than one subcarrier spacing. Of relevance in this paper, prior work \cite{cfo0}-\nocite{cfo1}\nocite{cfo2}\nocite{cfo3}\nocite{cfo4}\cite{cfo5} did not consider the impact of few-bit ADCs. Their developed synchronization pilots/sequences are therefore sensitive to the quantization distortion. This motivates us to construct new synchronization sequences that are robust to the low-resolution quantization.

In this paper, we propose and evaluate two novel frequency synchronization methods for downlink mmWave systems operating with low-resolution ADCs. The proposed two strategies exhibit different frequency synchronization performances under various configurations, and can be applied in different deployment scenarios. In our system model, the base station (BS) forms directional beams in the analog domain to send the synchronization signal towards the user equipment (UE). The UE employs a fully digital front end with low-resolution ADCs to detect the synchronization signal and conduct frequency synchronization. We focus on designing new synchronization sequences that are robust to the quantization distortion. We summarize the main contributions of the paper as follows:

\begin{itemize}
  \item \emph{New frequency synchronization sequences design}: We develop two double-sequence high-resolution CFO estimation methods for mmWave systems operating with low-resolution ADCs. In each method, we custom design two sequences (i.e., a sequence pair) for frequency synchronization. They are sent by the BS across two consecutive synchronization time-slots. We refer to the sequence pair as auxiliary sequences and sum-difference sequences in the proposed two methods. The key ingredient of the custom designed double-sequence structure (both auxiliary and sum-difference) is a ratio measure derived from the sequence pair, which is an invertible function of the CFO to be estimated. We use numerical examples to show that the ratio measures are robust to the quantization distortion brought by low-precision ADCs.
  \item \emph{Performance analysis of proposed low-resolution frequency synchronization methods}: We derive the Cramer-Rao lower bound (CRLB) of frequency estimation assuming $1$-bit ADCs. We show that the mean squared errors (MSEs) of our estimated CFOs using $1$-bit ADCs are close to the derived $1$-bit CRLB. Leveraging Bussgang's decomposition theorem \cite{buss2}, we derive the variance of the CFO estimates obtained via the proposed auxiliary sequences and sum-difference sequences based methods operating with low-resolution (e.g., $2$-$4$ bits) ADCs. These analytical results reveal that the CFO estimation performances highly depend on the double-sequence design parameters.
  \item \emph{Practical implementation of proposed low-resolution frequency synchronization methods}: Assuming multiple UEs, we formulate the corresponding low-resolution frequency synchronization problem as a min-max optimization problem. We first transform the min-max optimization problem into a minimization problem by exploiting certain long-term measurements and system statistics. We then solve the minimization problem by fine tuning the double-sequence design parameters such that the CFO estimation accuracy and the frequency range of estimation are jointly optimized. To better realize our proposed algorithms in practical systems, we implement additional signaling support and procedure at both the BS and UE sides.
\end{itemize}

We organize the rest of the paper as follows. In Section II, we specify the system and signal models for frequency synchronization in mmWave systems. In Section III, we present the design principle of the proposed auxiliary sequences and sum-difference sequences based CFO estimation strategies. We conduct performance analysis on our proposed methods in Section IV assuming few-bit ADCs. In Section V, we address several practical issues of implementing the proposed double-sequence low-resolution frequency synchronization designs in communications systems. We evaluate our proposed methods in Section VI assuming various channel models, quantization configurations and deployment scenarios. We draw our conclusions in Section VII.

\textbf{Notations}: $\bm{A}$ ($\textbf{\textsf{A}}$) is a matrix; $\bm{a}$ ($\textbf{\textsf{a}}$) is a vector; $a$ ($\textsf{a}$) is a scalar; $\textbf{\textsf{A}}$, $\textbf{\textsf{a}}$ and $\textsf{a}$ are Fourier transforms of $\bm{A}$, $\bm{a}$ and $a$; $|a|$ is the magnitude of the complex number
$a$; $(\cdot)^{\mathrm{T}}$ and $(\cdot)^{*}$ denote transpose and conjugate transpose; $((\cdot))_{N}$ represents the modulo-$N$ operation; $\mathrm{sign}(\cdot)$ extracts the sign of a real number; $\lfloor x\rfloor$ gives the largest integer less than or equal to $x$; $\angle(a)$ calculates the argument of the complex number $a$; $\left[\bm{A}\right]_{i,:}$ is the $i$-th row of $\bm{A}$; $\left[\bm{A}\right]_{:,j}$ is the $j$-th column of $\bm{A}$; $\left[\bm{A}\right]_{i,j}$ is the $(i,j)$-th entry of $\bm{A}$; $\left[\bm{a}\right]_{j}$ represents the $j$-th element of $\bm{a}$; $\bm{I}_{N}$ is the $N\times N$ identity matrix; $\mathcal{N}_{c}(\bm{a},\bm{A})$ is a complex Gaussian vector with mean $\bm{a}$ and covariance $\bm{A}$; $\Re\{\cdot\}$ and $\Im\{\cdot\}$ are used to extract the real and imaginary parts of given complex numbers; $\mathbb{E}[\cdot]$ is used to denote expectation; $\mathrm{diag}(\bm{a}^{\mathrm{T}})$ has $\big\{\big[\bm{a}\big]_{j}\big\}$ as its diagonal entries; and $\mathrm{diag}\big\{\big[\bm{a}\big]_{j}\big\}_{j=1}^{J}$ has $\big\{\big[\bm{a}\big]_{j},j=1,\cdots,J\big\}$ as its diagonal entries.

\section{System Model for Frequency Synchronization in MmWave Systems}
We consider a cellular system where synchronization and training data are sent periodically on directional beams, to support initial access \cite{ttfv,5gbmgnt}. Directional transmission is important in a mmWave system because it enables favorable received signal power to overcome higher pathloss and noise floor. This approach is used in the 3GPP 5G New Radio (NR) \cite{5gnr} and has been applied in other Wi-Fi standards \cite{11ayieee}. We assume that the UEs use fully digital front ends to detect the synchronization signal samples, which is realistic because there will only be a few antennas. In the following, we first present our employed system model, including the transceiver architecture, and antenna array configuration. We then develop the received synchronization signal model for our system and explain the conventional CFO estimation procedure using the ZC sequences.

\subsection{Transceiver architecture, array configuration and synchronization signal structure}
We consider a MIMO-OFDM system with $N$ subcarriers. The BS has $N_{\mathrm{tot}}$ transmit antennas. The UE has $M_{\mathrm{tot}}$ receive antennas. We further assume that the BS has $N_{\mathrm{RF}}$ radio frequency (RF) chains, and the UE uses $M_{\mathrm{RF}}$ RF chains. In this paper, we consider single-stream analog-only beamforming based directional synchronization at the BS (i.e., $N_{\mathrm{RF}}=1$) and fully digital baseband processing at the UE (i.e., $M_{\mathrm{tot}}=M_{\mathrm{RF}}$).

Due to their constant amplitude and zero autocorrelation \cite{popov}, ZC sequences are employed in the 3GPP LTE systems for downlink synchronization \cite{lte}. Denote the length of the employed ZC sequence by $N_{\mathrm{ZC}}$ and the sequence root index by $i$ ($i\in\left\{0,\cdots,N_{\mathrm{ZC}}-1\right\}$). For $m=0,\cdots,N_{\mathrm{ZC}}-1$, the sequence can be expressed as
\begin{equation}\label{zcfdd}
s_{i}[m] = \exp\left(-\mathrm{j}\frac{\pi m(m+1)i}{N_{\mathrm{ZC}}}\right).
\end{equation}
The cyclic auto-correlation of the ZC sequence results in a single dirac-impulse at zero-lag correlation:
\begin{equation}\label{fcorrex}
\chi[\upsilon] = \sum_{m=0}^{N_{\mathrm{ZC}}-1}s_{i}[m]s_{i}^{*}[((m+\upsilon))_{N_{\mathrm{ZC}}}]=\delta[\upsilon],\hspace{4mm}\upsilon=0,\cdots,N_{\mathrm{ZC}}-1.
\end{equation}
By exploiting (\ref{fcorrex}), the UE finds the estimate of the frame timing position that exhibits the largest peak in the correlation \cite{popov}. Besides the good correlation properties, an odd-length ZC sequence is symmetric with respect to its central element, i.e.,
\begin{equation}\label{zcsymmetric}
s_{i}[m'] = s_{i}[N_{\mathrm{ZC}}-1-m'],\hspace{2mm} m'=0,\cdots,\frac{N_{\mathrm{ZC}}-1}{2}.
\end{equation}
The UE can exploit this symmetry to perform the frequency synchronization, which will be elaborated in Section II-B.

Denote the frequency-domain modulated symbol on subcarrier $k=0,\cdots,N-1$ by $\textsf{d}[k]$. We map the ZC sequences to the central subcarriers as
\begin{eqnarray}\label{eqq1}
\textsf{d}[\lfloor(N-N_{\mathrm{ZC}}-1)/2\rfloor+m+1]=\Bigg\{\begin{array}{l}
            s_{i}[m],\hspace{2mm} m=0,\cdots,N_{\mathrm{ZC}}-1,\\
            0,\hspace{2mm}\textrm{otherwise},
          \end{array}
\end{eqnarray}
corresponding to the $N_{\mathrm{ZC}}$ subcarriers (out of $N$ subcarriers) surrounding the DC-carrier. In this paper, we configure the DC-carrier as zero as in the LTE systems \cite{lte}; note that in the 3GPP 5G NR systems (Release 15) \cite{5gnr}, no explicit DC-carrier is reserved for both the downlink and uplink.

\subsection{Received synchronization signal model}
To develop the received synchronization signal model, we assume: (i) a given UE $u\in\left\{1,\cdots,N_{\mathrm{UE}}\right\}$ in a single cell, where $N_{\mathrm{UE}}$ corresponds to the total number of active UEs in the cell of interest, and (ii) a given synchronization time-slot, which may be one OFDM symbol duration ($T_{\mathrm{s}}$). In Section VI, we simulate a more elaborate setting assuming multiple UEs and the frame structure adopted in the 3GPP LTE/NR.

The symbol vector $\textbf{\textsf{d}}$ in (\ref{eqq1}) is transformed to the time-domain via $N$-point IFFTs, generating the discrete-time samples $n=0,\cdots,N-1$ in one OFDM symbol duration as
\begin{equation}\label{tdzc}
d[n]=\frac{1}{\sqrt{N}}\sum_{k=0}^{N-1}\textsf{d}[k]e^{\mathrm{j}\frac{2\pi k}{N}n}.
\end{equation}
Before applying an $N_{\mathrm{tot}}\times 1$ wideband analog beamforming vector, a cyclic prefix (CP) is added to the symbol vector such that the length of the CP is greater than or equal to the maximum delay spread of the multi-path channels. Each sample in the symbol vector is then transmitted by a common wideband analog beamforming vector $\bm{f}$ from the BS, satisfying the power constraint $\left[\bm{f}\bm{f}^{*}\right]_{a,a}=\frac{1}{N_{\mathrm{tot}}}$, where $a=1,\cdots,N_{\mathrm{tot}}$.

Consider the $b$-th receive antenna ($b\in\left\{1,\cdots,M_{\mathrm{tot}}\right\})$ at UE $u$. After the timing synchronization and discarding the CP, the remaining time-domain received signal samples can be expressed as $\bm{q}_{u,b}=\big[q_{u,b}[0],\cdots,q_{u,b}[N-1]\big]^{\mathrm{T}}$. Denote the number of channel taps by $L_u$, the corresponding channel impulse response at tap $\ell\in\left\{0,\cdots,L_{u}-1\right\}$ by $\bm{H}_u[\ell]\in\mathbb{C}^{M_{\mathrm{tot}}\times N_{\mathrm{tot}}}$, and additive white Gaussian noise by $w_{u,b}[n]\sim \mathcal{N}_{c}(0,\sigma_u^{2})$. Denote the frequency mismatch with respect to the subcarrier spacing by $\varepsilon_u$. As the UE employs fully digital baseband processing, each receive antenna first quantizes the received synchronization signal with dedicated ADCs. Denote $\mathcal{Q}(\cdot)$ as the quantization function. For $n=0,\cdots,N-1$, the time-domain received signal samples are
\begin{eqnarray}\label{zcmap16}
q_{u,b}[n]=\mathcal{Q}\left(e^{\mathrm{j}\frac{2\pi\varepsilon_u}{N}n}\sum_{\ell=0}^{L_u-1}\left[\bm{H}_u[\ell]\right]_{b,:}\bm{f}d[n-\ell]+w_{u,b}[n]\right).
\end{eqnarray}
Across all receive antennas, UE $u$ selects the $\hat{b}$-th ($\hat{b}\in\left\{1,\cdots,M_{\mathrm{tot}}\right\}$) receive antenna that exhibits the highest received signal strength. By exploiting the symmetry of the employed ZC sequence and assuming perfect timing synchronization, the CFO can be estimated as
\begin{equation}\label{cfoestimation}
\hat{\varepsilon}_u=\frac{1}{\pi}\angle\left(\left[\sum_{n'=0}^{N/2-1}q_{u,\hat{b}}[n']d^{*}[n']\right]^{*}\left[\sum_{n''=N/2}^{N-1}q_{u,\hat{b}}[n'']d^{*}[n'']\right]\right).
\end{equation}
UE $u$ can then compensate the received signal samples with the estimated CFO as
\begin{equation}\label{cfocompensate}
\hat{q}_{u,b}[n]= e^{-\mathrm{j}\frac{2\pi\hat{\varepsilon}_u}{N}n}q_{u,b}[n].
\end{equation}
If $\mathcal{Q}(\cdot)$ in (\ref{zcmap16}) corresponds to low-resolution quantization (e.g., $1$-$4$ bits), the corresponding quantization distortion would damage the symmetry of the ZC sequence, leading to degraded CFO estimation performance. To see this, we apply Bussgang's theorem \cite{buss2} to first decouple the received synchronization signal into a useful signal part and an uncorrelated distortion component. For UE $u$, denote the normalized mean squared error (NMSE) of quantization by $\kappa_{u}$ \cite{jmps,cmollen}. The values of $\kappa_{u}\in[0,1]$ for various numbers of quantization bits are listed in \cite[Table~I]{chestjm}. Assuming independent and identically distributed (IID) Gaussian input to the quantizer, we rewrite (\ref{zcmap16}) as
\begin{figure}
\begin{center}
\subfigure[]{%
\includegraphics[width=.4\textwidth]{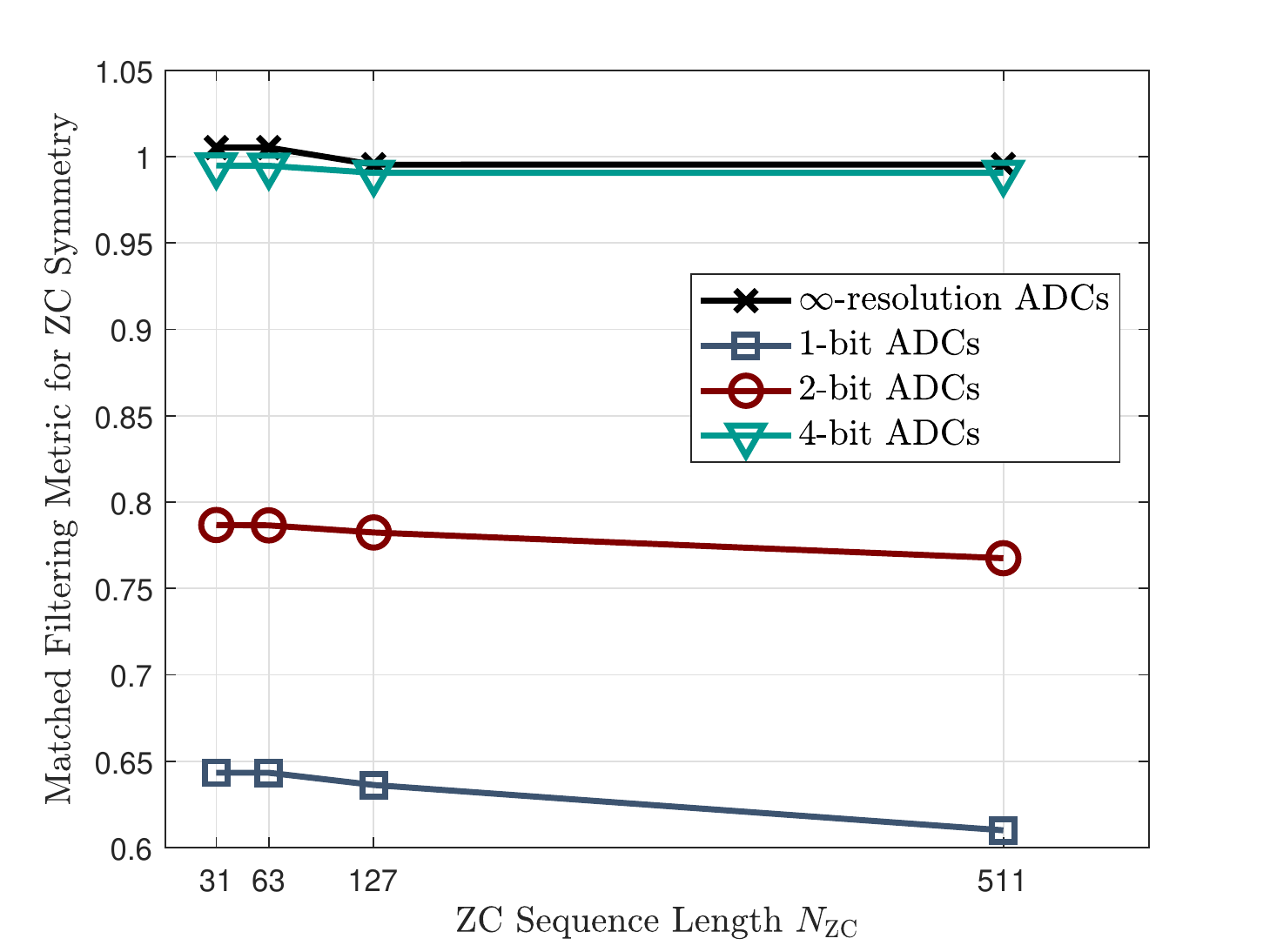}
\label{fig:simsubfigureZCsymmetrya}}
\hspace{-3.5mm}
\subfigure[]{%
\includegraphics[width=.4\textwidth]{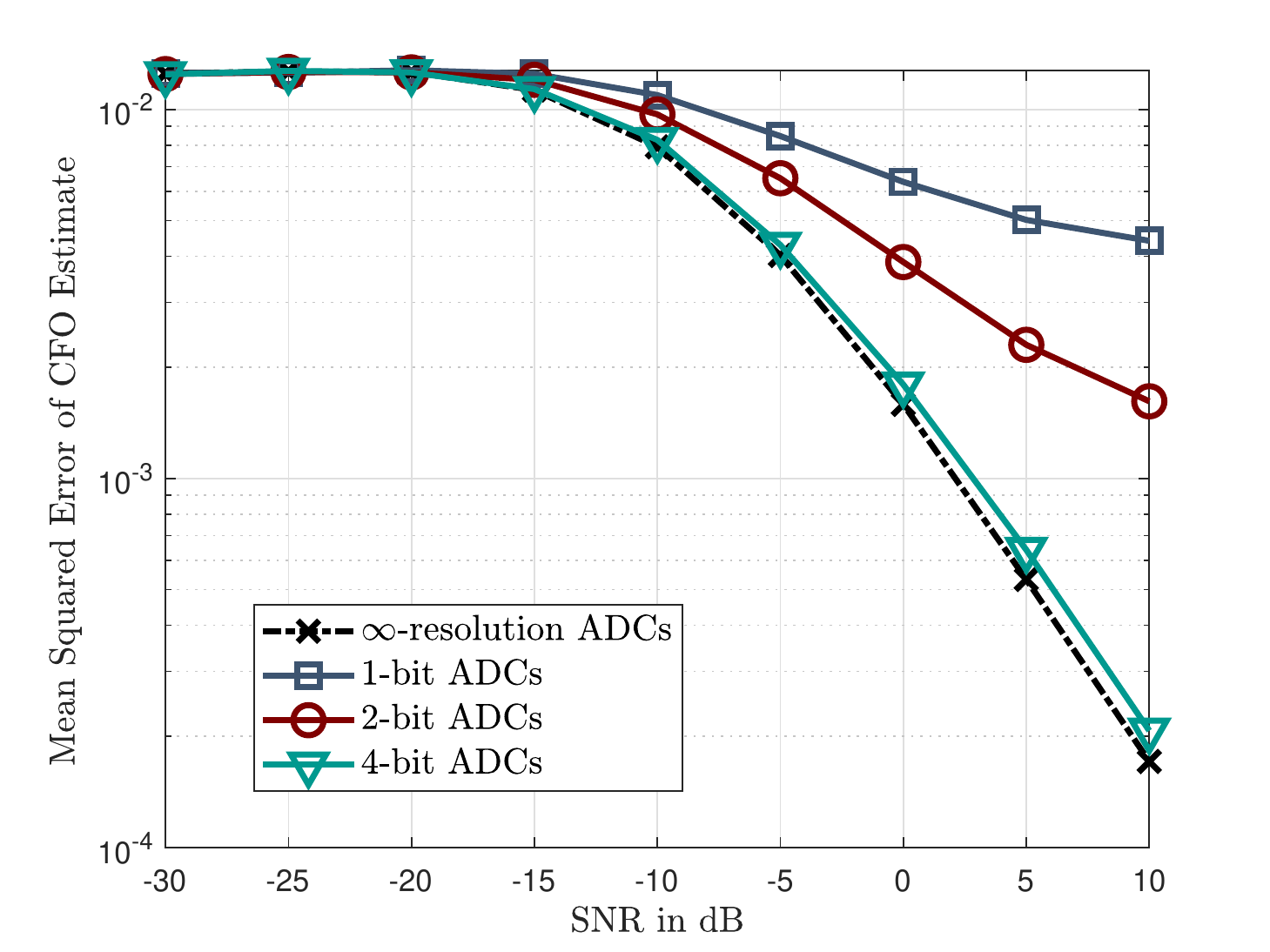}
\label{fig:simsubfigureZCsymmetryb}}
\caption{(a) Matched filtering metric for ZC symmetry versus the sequence length $N_{\mathrm{ZC}}$. AWGN channels are assumed with $10$~dB signal-to-noise ratio (SNR). The ADC resolutions are set as $1$, $2$ and $4$ bits. (b) Mean squared errors of the CFO estimates obtained via the ZC sequence based method under various SNRs. Single-path channels are assumed with a single UE. The ADC resolutions are set as $1$, $2$ and $4$ bits. The ZC sequence length is $63$ with the root index $25$.}
\label{fig:simfigureZCsymmetryab}
\end{center}
\end{figure}
\begin{eqnarray}\label{zcbuss}
q_{u,\hat{b}}[n]=(1-\kappa_{u})\Bigg(e^{\mathrm{j}\frac{2\pi\varepsilon_u}{N}n}\underbrace{\sum_{\ell=0}^{L_u-1}\left[\bm{H}_u[\ell]\right]_{\hat{b},:}\bm{f}d[n-\ell]}_{a_{u,\hat{b}}[n]}+w_{u,\hat{b}}[n]\Bigg)+v_{u,\hat{b}}[n],
\end{eqnarray}
where $\bm{v}_{u,\hat{b}}=\left[v_{u,\hat{b}}[0],\cdots,v_{u,\hat{b}}[N-1]\right]^{\mathrm{T}}$ is the quantization noise vector with covariance matrix $\sigma^{2}_{u,\mathcal{Q}}\bm{I}_{N}$, and here
\begin{equation}
\sigma^{2}_{u,\mathcal{Q}}=\kappa_{u}(1-\kappa_{u})\left(\mathbb{E}\left[\left|a_{u,\hat{b}}[n]\right|^{2}\right]+\sigma_u^2\right).
\end{equation}

Using (\ref{zcbuss}) to obtain $\hat{\varepsilon}_{u}$ via (\ref{cfoestimation}) may result in large CFO estimation errors because the symmetry of the ZC sequence could be significantly distorted by the low-resolution quantization. In Fig.~\ref{fig:simsubfigureZCsymmetrya}, we numerically characterize the impact of the low-resolution quantization on the symmetry of the ZC sequence. We first flip the second half of $\bm{q}_{u,\hat{b}}$ and denote
\begin{equation}\label{halfflip}
\breve{q}_{u,\hat{b}}[n'] = q_{u,\hat{b}}[N-1-n'],\hspace{2mm} n'=0,\cdots,\frac{N-1}{2}.
\end{equation}
We then define a matched filtering metric as
\begin{equation}
z=\mathbb{E}\Bigg[2\sum_{n'=0}^{N/2-1}q^{*}_{u,\hat{b}}[n']\breve{q}_{u,\hat{b}}[n']\Bigg].
\end{equation}
For $n'=0,\cdots,N/2-1$, if $q_{u,\hat{b}}[n']=\breve{q}_{u,\hat{b}}[n']$, i.e., perfect symmetry, the matched filtering metric $z$ achieves its peak, and corresponds to the zero-lag correlation value of the ZC sequence. It can be observed from Fig.~\ref{fig:simsubfigureZCsymmetrya} that with increase in the quantization distortion, the matched filtering metric significantly reduces, implying that the ZC symmetry is largely corrupted. In Fig.~\ref{fig:simsubfigureZCsymmetryb}, we evaluate the corresponding CFO estimation MSEs in single-path channels with a single UE. Assuming $1$-bit and $2$-bit ADCs, we observe that the frequency synchronization performances are much poorer than that of the infinite-resolution ADCs. We conclude from Fig.~\ref{fig:simfigureZCsymmetryab} that to achieve promising low-resolution frequency synchronization performance, it becomes desirable to develop new synchronization sequences such that the corresponding metrics used for estimating the CFO are robust to the quantization distortion.
\begin{figure}
\centering
\includegraphics[width=6.15in]{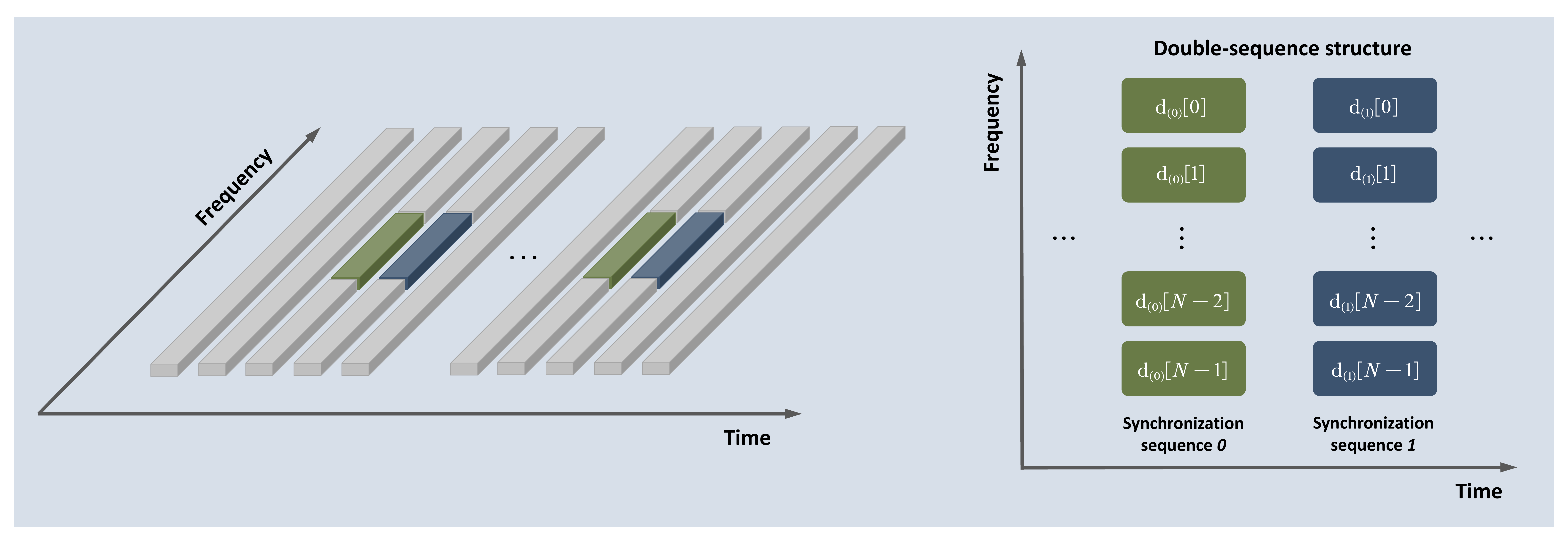}
\caption{Time-frequency resource mapping of the proposed double-sequence frequency synchronization. Synchronization sequences $0$ and $1$ in the double-sequence structure are transmitted by the BS across two consecutive synchronization time-slots.}
\label{fig:figureDoubleSeq}
\end{figure}
\section{Proposed Double-Sequence Frequency Synchronization}
We propose two novel double-sequence frequency synchronization methods. In this section, we explicitly explain the design principle of the proposed methods assuming infinite-resolution quantization. In later sections, we will illustrate the detailed implementation procedure of our proposed methods along with comprehensive performance evaluations assuming low-resolution quantization. In Fig.~\ref{fig:figureDoubleSeq}, we depict the basic double-sequence structure and its time-frequency resource mapping. As can be seen from the right-hand side of Fig.~\ref{fig:figureDoubleSeq}, two length-$N$ sequences are periodically transmitted across two consecutive synchronization time-slots in a time-division multiplexing (TDM) manner.

\subsection{Auxiliary sequences based frequency synchronization}
We denote the time-domain samples of the two sequences by $\bm{d}_{0}=\big[d_{0}[0],\cdots,d_{0}[N-1]\big]^{\mathrm{T}}$ and $\bm{d}_{1}=\left[d_{1}[0],\cdots,d_{1}[N-1]\right]^{\mathrm{T}}$. Further, we construct $\bm{d}_{0}$ and $\bm{d}_{1}$ as
\begin{eqnarray}
\bm{d}_{0}&=&\left[d_{0}[0],d_{0}[1],\cdots,d_{0}[N-1]\right]^{\mathrm{T}}=\left[1,e^{-\mathrm{j}(\theta-\delta)},\cdots,e^{-\mathrm{j}(N-1)(\theta-\delta)}\right]^{\mathrm{T}}\label{axvia0}\\
\bm{d}_{1}&=&\left[d_{1}[0],d_{1}[1],\cdots,d_{1}[N-1]\right]^{\mathrm{T}}=\left[1,e^{-\mathrm{j}(\theta+\delta)},\cdots,e^{-\mathrm{j}(N-1)(\theta+\delta)}\right]^{\mathrm{T}}\label{axvia1},
\end{eqnarray}
and they are sent by the BS using two consecutive synchronization time-slots, say, synchronization time-slots $0$ and $1$. For $k=0,\cdots,N-1$, the frequency-domain samples that correspond to $\bm{d}_{0}$ and $\bm{d}_{1}$ are
\begin{equation}\label{fftrelt}
\textsf{d}_{0}[k]=\frac{1}{\sqrt{N}}\sum_{n=0}^{N-1}d_{0}[n]e^{-\mathrm{j}\frac{2\pi n}{N}k},\hspace{3mm}\textsf{d}_{1}[k]=\frac{1}{\sqrt{N}}\sum_{n=0}^{N-1}d_{1}[n]e^{-\mathrm{j}\frac{2\pi n}{N}k}.
\end{equation}

For synchronization time-slot $0$ and $n=0,\cdots,N-1$, we express the corresponding time-domain received signal samples as (similar to (\ref{zcmap16}))
\begin{equation}\label{ax0ori}
q^{0}_{u,b}[n]=\mathcal{Q}\left(e^{\mathrm{j}\frac{2\pi\varepsilon_u}{N}n}\sum_{\ell=0}^{L_u-1}\left[\bm{H}_u[\ell]\right]_{b,:}\bm{f}d_{0}[n-\ell]+w^{0}_{u,b}[n]\right).
\end{equation}
Assuming infinite-resolution quantization and neglecting noise,
\begin{eqnarray}\label{ax0}
q^{0}_{u,b}[n]=e^{\mathrm{j}\frac{2\pi\varepsilon_u}{N}n}\sum_{\ell=0}^{L_u-1}\left[\bm{H}_u[\ell]\right]_{b,:}\bm{f}d_{0}[n-\ell].
\end{eqnarray}
Defining the normalized CFO for UE $u$ as $\mu_{u}=2\pi\varepsilon_{u}/N$,
\begin{eqnarray}\label{ax0fur}
q^{0}_{u,b}[n]&=&e^{\mathrm{j}\mu_{u}n}\sum_{\ell=0}^{L_u-1}\left[\bm{H}_u[\ell]\right]_{b,:}\bm{f}d_{0}[n-\ell]\\
&=&e^{\mathrm{j}\mu_{u}n}\sum_{\ell=0}^{L_u-1}\left[\bm{H}_u[\ell]\right]_{b,:}\bm{f}e^{-\mathrm{j}(n-\ell)(\theta-\delta)}\\
&=&e^{\mathrm{j}\mu_{u}n}e^{-\mathrm{j}n(\theta-\delta)}\sum_{\ell=0}^{L_u-1}\left[\bm{H}_u[\ell]\right]_{b,:}\bm{f}e^{\mathrm{j}\ell(\theta-\delta)}\\
&=&e^{\mathrm{j}n(\mu_{u}-\theta+\delta)}\left[\textbf{\textsf{H}}_{u}\left(e^{-\mathrm{j}(\theta-\delta)}\right)\right]_{b,:}\bm{f}.
\end{eqnarray}
Denote the selected receive antenna index by $\hat{b}$. Using the received signal samples $\bm{q}^{0}_{u,\hat{b}}=\left[q^{0}_{u,\hat{b}}[0],\cdots,q^{0}_{u,\hat{b}}[N-1]\right]^{\mathrm{T}}$ from synchronization time-slot $0$, UE $u$ calculates
\begin{eqnarray}
p^{0}_{u,\hat{b}}&=&\left(\sum_{n=0}^{N-1}q^{0}_{u,\hat{b}}[n]\right)\left(\sum_{n=0}^{N-1}q^{0}_{u,\hat{b}}[n]\right)^{*}\label{testarray}\\
&=&\left|\left[\textbf{\textsf{H}}_{u}\left(e^{-\mathrm{j}(\theta-\delta)}\right)\right]_{\hat{b},:}\bm{f}\right|^{2}\left|\sum_{n=0}^{N-1}e^{\mathrm{j}n(\mu_u-\theta+\delta)}\right|^{2}\\
&\stackrel{(\star)}{=}&\left|\left[\textbf{\textsf{H}}_{u}\left(e^{-\mathrm{j}(\theta-\delta)}\right)\right]_{\hat{b},:}\bm{f}\right|^{2}\frac{\sin^{2}\left(\frac{N(\mu_u-\theta+\delta)}{2}\right)}{\sin^{2}\left(\frac{\mu_u-\theta+\delta}{2}\right)},\label{axss0}
\end{eqnarray}
where ($\star$) is obtained via $\left|\sum_{m=1}^{M}e^{\mathrm{j}(m-1)x}\right|^{2}=\frac{\sin^{2}\left(\frac{Mx}{2}\right)}{\sin^{2}\left(\frac{x}{2}\right)}$. Similarly, using the received signal samples $\bm{q}^{1}_{u,\hat{b}}=\left[q^{1}_{u,\hat{b}}[0],\cdots,q^{1}_{u,\hat{b}}[N-1]\right]^{\mathrm{T}}$ from synchronization time-slot $1$, UE $u$ computes
\begin{eqnarray}
p^{1}_{u,\hat{b}}&=&\left(\sum_{n=0}^{N-1}q^{1}_{u,\hat{b}}[n]\right)\left(\sum_{n=0}^{N-1}q^{1}_{u,\hat{b}}[n]\right)^{*}\\
&=&\left|\left[\textbf{\textsf{H}}_{u}\left(e^{-\mathrm{j}(\theta+\delta)}\right)\right]_{\hat{b},:}\bm{f}\right|^{2}\frac{\sin^{2}\left(\frac{N(\mu_u-\theta-\delta)}{2}\right)}{\sin^{2}\left(\frac{\mu_u-\theta-\delta}{2}\right)}.\label{axss1}
\end{eqnarray}
By letting $\delta = 2k'\pi/N$ ($k'=1,\cdots,\frac{N}{4}$), we can rewrite (\ref{axss0}) and (\ref{axss1}) as
\begin{equation}\label{ccomb}
p^{0}_{u,\hat{b}}=\left|\left[\textbf{\textsf{H}}_{u}\left(e^{-\mathrm{j}(\theta-\delta)}\right)\right]_{\hat{b},:}\bm{f}\right|^{2}\frac{\sin^{2}\left(\frac{N(\mu_u-\theta)}{2}\right)}{\sin^{2}\left(\frac{\mu_u-\theta+\delta}{2}\right)},\hspace{1.5mm}p^{1}_{u,\hat{b}}=\left|\left[\textbf{\textsf{H}}_{u}\left(e^{-\mathrm{j}(\theta+\delta)}\right)\right]_{\hat{b},:}\bm{f}\right|^{2}\frac{\sin^{2}\left(\frac{N(\mu_u-\theta)}{2}\right)}{\sin^{2}\left(\frac{\mu_u-\theta-\delta}{2}\right)}.
\end{equation}
Using $p^{0}_{u,\hat{b}}$ and $p^{1}_{u,\hat{b}}$ in (\ref{ccomb}), UE $u$ calculates a ratio metric as
\begin{eqnarray}
\alpha_{u} &=& \frac{p^{0}_{u,\hat{b}}-p^{1}_{u,\hat{b}}}{p^{0}_{u,\hat{b}}+p^{1}_{u,\hat{b}}}\label{alpharatiometric}\\
&=& \frac{\sin^{2}\left(\frac{\mu_u-\theta-\delta}{2}\right)-\sin^{2}\left(\frac{\mu_u-\theta+\delta}{2}\right)}{\sin^{2}\left(\frac{\mu_u-\theta-\delta}{2}\right)+\sin^{2}\left(\frac{\mu_u-\theta+\delta}{2}\right)}\\
&=& -\frac{\sin(\mu_u-\theta)\sin(\delta)}{1-\cos(\mu_u-\theta)\cos(\delta)}\label{howtoinvertr},
\end{eqnarray}
which does not depend on the selected receive antenna index $\hat{b}$. According to \cite[Lemma~1]{dztrans}, if $|\mu_u-\theta|<\delta$, the ratio measure $\alpha_u$ is a monotonic decreasing function of $\mu_u-\theta$ and invertible with respect to $\mu_{u}-\theta$. Via the inverse function, we can derive the estimated value of $\mu_u$ as
\begin{equation}\label{axratiometricth}
\hat{\mu}_{u}=\theta-\arcsin\left(\frac{\alpha_u\sin(\delta)-\alpha_u\sqrt{1-\alpha^2_u}\sin(\delta)\cos(\delta)}{\sin^{2}(\delta)+\alpha^{2}_{u}\cos^{2}(\delta)}\right).
\end{equation}
We can then obtain the super-resolution CFO estimate for UE $u$ as $\hat{\varepsilon}_{u}=\frac{N}{2\pi}\hat{\mu}_{u}$, comprising both the integer and fractional components. Note that if $\alpha_u$ is perfect, i.e., not impaired by noise and quantization distortion, the CFO can be perfectly recovered, i.e., $\hat{\varepsilon}_{u}=\varepsilon_u$.

\subsection{Sum-difference sequences based frequency synchronization}
In the proposed sum-difference sequences based approach, the time-domain synchronization sequences $\bm{d}_{\Sigma}$ and $\bm{d}_{\Delta}$ exhibit different forms from $\bm{d}_0$ and $\bm{d}_{1}$ in (\ref{axvia0}) and (\ref{axvia1}). Specifically, we construct $\bm{d}_{\Sigma}$ and $\bm{d}_{\Delta}$ as
\begin{eqnarray}
\bm{d}_{\Sigma}&=&\left[d_{\Sigma}[0],\cdots,d_{\Sigma}[N/2-1],d_{\Sigma}[N/2],\cdots,d_{\Sigma}[N-1]\right]^{\mathrm{T}}\nonumber\\
&=&\left[1,\cdots,e^{-\mathrm{j}(N/2-1)\eta},e^{-\mathrm{j}(N/2)\eta},\cdots,e^{-\mathrm{j}(N-1)\eta}\right]^{\mathrm{T}}\label{sdvia0}\\
\bm{d}_{\Delta}&=&\left[d_{\Delta}[0],\cdots,d_{\Delta}[N/2-1],d_{\Delta}[N/2],\cdots,d_{\Delta}[N-1]\right]^{\mathrm{T}}\nonumber\\
&=&\left[1,\cdots,e^{-\mathrm{j}(N/2-1)\eta},-e^{-\mathrm{j}(N/2)\eta},\cdots,-e^{-\mathrm{j}(N-1)\eta}\right]^{\mathrm{T}}\label{sdvia1},
\end{eqnarray}
which are referred to as sum and difference synchronization sequences. They are transmitted via two consecutive synchronization time-slots. Their frequency-domain counterparts $\textbf{\textsf{d}}_{\Sigma}$ and $\textbf{\textsf{d}}_{\Delta}$ can be similarly obtained following (\ref{fftrelt}). Note that the first halves of the sum and difference synchronization sequences $\bm{d}_{\Sigma}$ and $\bm{d}_{\Delta}$ are identical, while the second half of $\bm{d}_{\Delta}$ is the additive inverse of the second half of $\bm{d}_{\Sigma}$.

For synchronization time-slot $0$, and therefore the corresponding sum synchronization sequence $\bm{d}_{\Sigma}$, we express the time-domain received signal samples as ($n=0,\cdots,N-1$)
\begin{equation}\label{sdseqdeng}
q^{\Sigma}_{u,b}[n]=\mathcal{Q}\left(e^{\mathrm{j}\frac{2\pi\varepsilon_u}{N}n}\sum_{\ell=0}^{L_u-1}\left[\bm{H}_u[\ell]\right]_{b,:}\bm{f}d_{\Sigma}[n-\ell]+w^{\Sigma}_{u,b}[n]\right).
\end{equation}
Neglecting noise and assuming $\mathcal{Q}(\cdot)$ as the infinite-resolution quantization function,
\begin{eqnarray}
q^{\Sigma}_{u,b}[n]&=&e^{\mathrm{j}\frac{2\pi\varepsilon_u}{N}n}\sum_{\ell=0}^{L_u-1}\left[\bm{H}_u[\ell]\right]_{b,:}\bm{f}d_{\Sigma}[n-\ell]\\
&=&e^{\mathrm{j}\mu_{u}n}\sum_{\ell=0}^{L_u-1}\left[\bm{H}_u[\ell]\right]_{b,:}\bm{f}d_{\Sigma}[n-\ell]\\
&=&e^{\mathrm{j}\mu_{u}n}\sum_{\ell=0}^{L_u-1}\left[\bm{H}_u[\ell]\right]_{b,:}\bm{f}e^{-\mathrm{j}(n-\ell)\eta}\\
&=&e^{\mathrm{j}n(\mu_{u}-\eta)}\sum_{\ell=0}^{L_u-1}\left[\bm{H}_u[\ell]\right]_{b,:}\bm{f}e^{\mathrm{j}\ell\eta}\\
&=&e^{\mathrm{j}n(\mu_{u}-\eta)}\left[\textbf{\textsf{H}}_{u}\left(e^{-\mathrm{j}\eta}\right)\right]_{b,:}\bm{f}.
\end{eqnarray}
Assuming $\hat{b}$ as the selected receive antenna index for UE $u$, we sum $\bm{q}^{\Sigma}_{u,\hat{b}}=\Big[q^{\Sigma}_{u,\hat{b}}[0],\cdots,q^{\Sigma}_{u,\hat{b}}[N-1]\Big]^{\mathrm{T}}$ over all samples and obtain
\begin{eqnarray}
p^{\Sigma}_{u,\hat{b}} &=& \sum_{n=0}^{N-1}q^{\Sigma}_{u,\hat{b}}[n]\\
&=& \left[\textbf{\textsf{H}}_{u}\left(e^{-\mathrm{j}\eta}\right)\right]_{\hat{b},:}\bm{f}\sum_{n=0}^{N-1}e^{\mathrm{j}n(\mu_{u}-\eta)}\\
&=& \left[\textbf{\textsf{H}}_{u}\left(e^{-\mathrm{j}\eta}\right)\right]_{\hat{b},:}\bm{f}\left(1+e^{\mathrm{j}(N/2)(\mu_{u}-\eta)}\right)\sum_{n'=0}^{N/2-1}e^{\mathrm{j}n'(\mu_u-\eta)}\\
&=& \left[\textbf{\textsf{H}}_{u}\left(e^{-\mathrm{j}\eta}\right)\right]_{\hat{b},:}\bm{f}\left(1+e^{\mathrm{j}(N/2)(\mu_{u}-\eta)}\right)e^{\mathrm{j}\frac{N/2-1}{2}(\mu_u-\eta)}\frac{\sin\left(\frac{N/2}{2}(\mu_{u}-\eta)\right)}{\sin\left((\mu_u-\eta)/2\right)}.\label{sumsyn0}
\end{eqnarray}
Similar to (\ref{sumsyn0}), for synchronization time-slot $1$, we have
\begin{eqnarray}
p^{\Delta}_{u,\hat{b}}&=& \left[\textbf{\textsf{H}}_{u}\left(e^{-\mathrm{j}\eta}\right)\right]_{\hat{b},:}\bm{f}\left(1-e^{\mathrm{j}(N/2)(\mu_{u}-\eta)}\right)\sum_{n'=0}^{N/2-1}e^{\mathrm{j}n'(\mu_u-\eta)}\\
&=& \left[\textbf{\textsf{H}}_{u}\left(e^{-\mathrm{j}\eta}\right)\right]_{\hat{b},:}\bm{f}\left(1-e^{\mathrm{j}(N/2)(\mu_{u}-\eta)}\right)e^{\mathrm{j}\frac{N/2-1}{2}(\mu_u-\eta)}\frac{\sin\left(\frac{N/2}{2}(\mu_{u}-\eta)\right)}{\sin\left((\mu_u-\eta)/2\right)}.\label{diffsyn0}
\end{eqnarray}
Using $p^{\Sigma}_{u,\hat{b}}$ and $p^{\Delta}_{u,\hat{b}}$, UE $u$ calculates a ratio measure as
\begin{eqnarray}
\beta_u &=& \Im\left\{\frac{p^{\Sigma}_{u,\hat{b}}}{p^{\Delta}_{u,\hat{b}}}\right\}=\frac{\sin\left(\frac{N}{2}(\mu_{u}-\eta)\right)}{1-\cos\left(\frac{N}{2}(\mu_u-\eta)\right)}\label{betaratiometric}\\
&=&\cot\left(\frac{N}{4}(\mu_{u}-\eta)\right),
\end{eqnarray}
which does not depend on the selected receive antenna index $\hat{b}$. By exploiting some trigonometric identities, for $\mu_{u}-\eta\in\big(4i\pi/N,4(i+1)\pi/N\big)$ with $i\in\mathbb{Z}$, we can invert the ratio measure $\beta_u$ and obtain
\begin{equation}\label{sdratiometricth}
\hat{\mu}_{u}=\eta+\frac{4}{N}\cot^{-1}(\beta_u),
\end{equation}
which is then used by UE $u$ to compute the CFO estimate as $\hat{\varepsilon}_u=\frac{N}{2\pi}\hat{\mu}_{u}$. If $\beta_u$ is perfect, i.e., not impaired by noise and quantization distortion, the CFO can be perfectly recovered, i.e., $\varepsilon_u=\hat{\varepsilon}_u$. As $\mu_u$ ($\hat{\mu}_{u}$) and $\varepsilon_u$ ($\hat{\varepsilon}_u$) are equivalent, we use the normalized CFO $\mu_u$ and its estimate $\hat{\mu}_{u}$ throughout the rest of this paper unless otherwise specified.

If $\mathcal{Q}(\cdot)$ in (\ref{ax0ori}) and (\ref{sdseqdeng}) corresponds to the low-resolution quantization function (e.g., $1$-$4$ bits), the quantization distortion may damage the monotonic properties of the ratio metrics, resulting in increased CFO estimation errors. In Sections V and VI, we use both analytical and numerical examples to show that our proposed approaches are robust to the low-resolution quantization; we further show that the proposed two design options exhibit different frequency synchronization performances under different settings (propagation condition, channel variation and etc.). Different cells may flexibly configure their employed synchronization sequences (either auxiliary or sum-difference) for frequency synchronization, depending on their own demands.

\section{Low-Resolution Double-Sequence Frequency Synchronization}
In Section IV-A, we first present the CRLB for frequency estimation under $1$-bit quantization. We use numerical examples to show that under $1$-bit quantization, the mean squared errors of our estimated CFOs are close to the derived $1$-bit CRLB. By exploiting Bussgang's theorem, we then derive the variance of the CFO estimates obtained via the proposed strategies assuming finite low-resolution (e.g., $2$-$4$ bits) quantization. We further examine the robustness of our proposed methods to the quantization distortion. Leveraging the variance results derived in Section IV-B and IV-C, we provide design insights of implementing our proposed algorithms in practical cellular systems.
\subsection{Cramer-Rao lower bound for frequency estimation under 1-bit ADCs}
We first denote $\bm{d}_{(0)}$ and $\bm{d}_{(1)}$ as the two sequences transmitted during synchronization time-slots $0$ and $1$. They correspond to either auxiliary sequences or sum-difference sequences in our proposed methods. Using $\bm{d}_{(0)}$ and $\bm{d}_{(1)}$ and assuming the $b$-th receive antenna at UE $u$, we present a generic received signal model for the proposed double-sequence structure as ($n=0,\cdots,N-1$)
\begin{eqnarray}
q^{(0)}_{u,b}[n]&=&\mathcal{Q}\Bigg(e^{\mathrm{j}\frac{2\pi\varepsilon_u}{N}n}\underbrace{\sum_{\ell=0}^{L_u-1}\left[\bm{H}_u[\ell]\right]_{b,:}\bm{f}d_{(0)}[n-\ell]}_{a^{(0)}_{u,b}[n]}+w^{(0)}_{u,b}[n]\Bigg)\label{vectorize0}\\
q^{(1)}_{u,b}[n]&=&\mathcal{Q}\Bigg(e^{\mathrm{j}\frac{2\pi\varepsilon_u}{N}n}\underbrace{\sum_{\ell=0}^{L_u-1}\left[\bm{H}_u[\ell]\right]_{b,:}\bm{f}d_{(1)}[n-\ell]}_{a^{(1)}_{u,b}[n]}+w^{(1)}_{u,b}[n]\Bigg)\label{vectorize1}.
\end{eqnarray}
Denote $\bm{A}_{u,b}^{(0)} = \mathrm{diag}\Big(\Big[a_{u,b}^{(0)}[0],\cdots,a_{u,b}^{(0)}[N-1]\Big]^{\mathrm{T}}\Big)$ and $\bm{A}_{u,b}^{(1)} = \mathrm{diag}\Big(\Big[a_{u,b}^{(1)}[0],\cdots,a_{u,b}^{(1)}[N-1]\Big]^{\mathrm{T}}\Big)$. We further define $\bm{e}_u=\left[1,e^{\mathrm{j}\frac{2\pi\varepsilon_u}{N}},\cdots,e^{\mathrm{j}\frac{2\pi\varepsilon_u}{N}(N-1)}\right]^{\mathrm{T}}$. For all samples $n=0,\cdots,N-1$, we write (\ref{vectorize0}) and (\ref{vectorize1}) in vector forms as
\begin{eqnarray}\label{defbisheng}
\bm{q}_{u,b}^{(0)}=\mathcal{Q}\left(\bm{A}_{u,b}^{(0)}\bm{e}_{u}+\bm{w}_{u,b}^{(0)}\right),\hspace{3mm}\bm{q}_{u,b}^{(1)}=\mathcal{Q}\left(\bm{A}_{u,b}^{(1)}\bm{e}_{u}+\bm{w}_{u,b}^{(1)}\right).
\end{eqnarray}
To compute the CRLB for $1$-bit quantization, we first define
\begin{eqnarray}
\widetilde{\bm{q}}_{u,b}^{(0)}=\left(
                             \begin{array}{c}
                               \Re\big\{\bm{q}_{u,b}^{(0)}\big\} \\
                               \Im\big\{\bm{q}_{u,b}^{(0)}\big\} \\
                             \end{array}
                           \right),\hspace{3mm}\widetilde{\bm{q}}_{u,b}^{(1)}=\left(
                             \begin{array}{c}
                               \Re\big\{\bm{q}_{u,b}^{(1)}\big\} \\
                               \Im\big\{\bm{q}_{u,b}^{(1)}\big\} \\
                             \end{array}
                           \right),\hspace{3mm}\widetilde{\bm{e}}_{u}=\left(
                             \begin{array}{c}
                               \Re\left\{\bm{e}_{u}\right\} \\
                               \Im\left\{\bm{e}_{u}\right\} \\
                             \end{array}
                           \right),
\end{eqnarray}
\begin{eqnarray}
                           \hspace{1mm}\widetilde{\bm{w}}_{u,b}^{(0)}=\left(
                             \begin{array}{c}
                               \Re\big\{\bm{w}_{u,b}^{(0)}\big\} \\
                               \Im\big\{\bm{w}_{u,b}^{(0)}\big\} \\
                             \end{array}
                           \right),\hspace{3mm}\widetilde{\bm{w}}_{u,b}^{(1)}=\left(
                             \begin{array}{c}
                               \Re\big\{\bm{w}_{u,b}^{(1)}\big\} \\
                               \Im\big\{\bm{w}_{u,b}^{(1)}\big\} \\
                             \end{array}
                           \right),
\end{eqnarray}
and
\begin{eqnarray}
\widetilde{\bm{A}}_{u,b}^{(0)}=\left(
                                 \begin{array}{cc}
                                   \Re\Big\{\bm{A}_{u,b}^{(0)}\Big\} & -\Im\Big\{\bm{A}_{u,b}^{(0)}\Big\} \\
                                   \Im\Big\{\bm{A}_{u,b}^{(0)}\Big\} & \Re\Big\{\bm{A}_{u,b}^{(0)}\Big\} \\
                                 \end{array}
                               \right),\hspace{3mm}\widetilde{\bm{A}}_{u,b}^{(1)}=\left(
                                 \begin{array}{cc}
                                   \Re\Big\{\bm{A}_{u,b}^{(1)}\Big\} & -\Im\Big\{\bm{A}_{u,b}^{(1)}\Big\} \\
                                   \Im\Big\{\bm{A}_{u,b}^{(1)}\Big\} & \Re\Big\{\bm{A}_{u,b}^{(1)}\Big\} \\
                                 \end{array}
                               \right).
\end{eqnarray}
\begin{figure}
\begin{center}
\subfigure[]{%
\includegraphics[width=.4\textwidth]{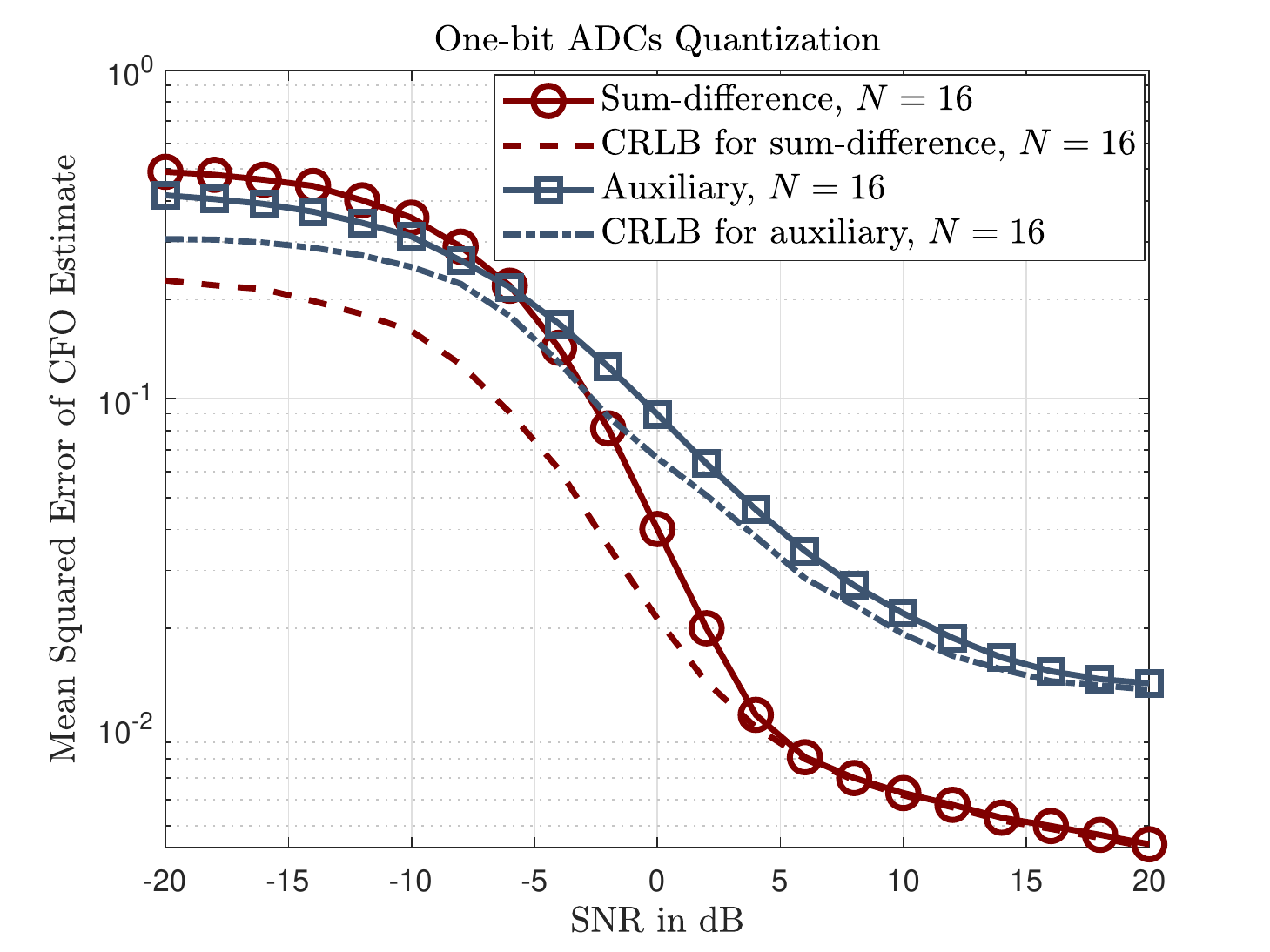}
\label{fig:crlbsubfigure1a}}
\hspace{-3.5mm}
\subfigure[]{%
\includegraphics[width=.4\textwidth]{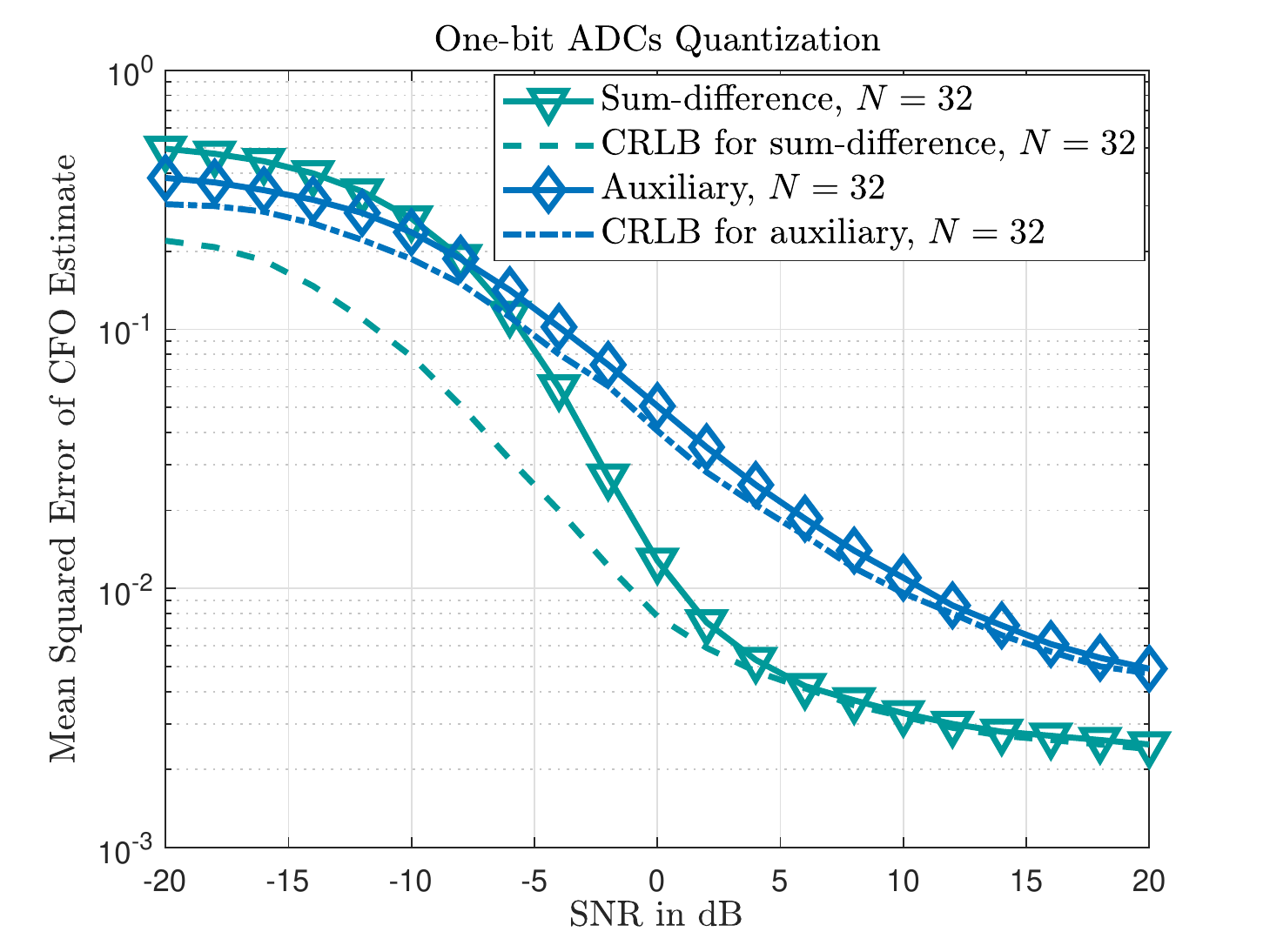}
\label{fig:crlbsubfigure1b}}
\caption{CRLBs and MSEs of the CFO estimates under $1$-bit ADCs quantization. Both the proposed auxiliary sequences and sum-difference sequences based strategies are evaluated. Single-path channels are assumed with a single UE. $N_{\mathrm{tot}}=16$ and $M_{\mathrm{tot}}=8$. (a) $N=16$. (b) $N=32$.}
\label{fig:crlbfigure1ab}
\end{center}
\end{figure}

We can rewrite (\ref{defbisheng}) in real-valued form as
\begin{eqnarray}\label{defbishengReal}
\widetilde{\bm{q}}_{u,b}^{(0)}=\mathcal{Q}\left(\widetilde{\bm{A}}_{u,b}^{(0)}\widetilde{\bm{e}}_{u}+\widetilde{\bm{w}}_{u,b}^{(0)}\right),\hspace{3mm}\widetilde{\bm{q}}_{u,b}^{(1)}=\mathcal{Q}\left(\widetilde{\bm{A}}_{u,b}^{(1)}\widetilde{\bm{e}}_{u}+\widetilde{\bm{w}}_{u,b}^{(1)}\right).
\end{eqnarray}
Denote the covariance matrix of $\widetilde{\bm{w}}_{u,b}^{(0)}$ ($\widetilde{\bm{w}}_{u,b}^{(1)}$) as $\widetilde{\sigma}_{u}^{2}\bm{I}_{2N}$ with $\widetilde{\sigma}^{2}_{u}=\sigma^{2}_{u}/2$, and
\begin{eqnarray}
\widetilde{\bm{g}}_{u,b}^{(0)} = \frac{1}{\widetilde{\sigma}_u}\sum_{q=1}^{2N}\left[\widetilde{\bm{A}}_{u,b}^{(0)}\right]_{:,q}\left[\widetilde{\bm{e}}_{u}\right]_{q},\hspace{3mm}\widetilde{\bm{g}}_{u,b}^{(1)} = \frac{1}{\widetilde{\sigma}_u}\sum_{q=1}^{2N}\left[\widetilde{\bm{A}}_{u,b}^{(1)}\right]_{:,q}\left[\widetilde{\bm{e}}_{u}\right]_{q}.
\end{eqnarray}
Assuming $1$-bit element-wise quantization such that
\begin{equation}\label{1bitqfunc}
\mathcal{Q}(x) = \mathrm{sign}(\Re\{x\})+\mathrm{j}\cdot\mathrm{sign}(\Im\{x\}),
\end{equation}
we can then obtain the real-valued Fisher information matrix as
\begin{eqnarray}
\widetilde{\bm{I}}_{\widetilde{\bm{e}}_{u},b}&=&\left(\widetilde{\bm{A}}_{u,b}^{(0)}\right)^{\mathrm{T}}\mathrm{diag}\left\{\frac{1}{\widetilde{\sigma}^{2}_u}\frac{\phi^{2}\left(\left[\widetilde{\bm{g}}_{u,b}^{(0)}\right]_{p}\right)}{\Phi\left(\left[\widetilde{\bm{g}}_{u,b}^{(0)}\right]_{p}\right)\left[1-\Phi\left(\left[\widetilde{\bm{g}}_{u,b}^{(0)}\right]_{p}\right)\right]}\right\}_{p=1}^{2N}\widetilde{\bm{A}}_{u,b}^{(0)}\nonumber\\
&+&\left(\widetilde{\bm{A}}_{u,b}^{(1)}\right)^{\mathrm{T}}\mathrm{diag}\left\{\frac{1}{\widetilde{\sigma}^{2}_u}\frac{\phi^{2}\left(\left[\widetilde{\bm{g}}_{u,b}^{(1)}\right]_{p}\right)}{\Phi\left(\left[\widetilde{\bm{g}}_{u,b}^{(1)}\right]_{p}\right)\left[1-\Phi\left(\left[\widetilde{\bm{g}}_{u,b}^{(1)}\right]_{p}\right)\right]}\right\}_{p=1}^{2N}\widetilde{\bm{A}}_{u,b}^{(1)},
\end{eqnarray}
where $\Phi(x)=\int_{-\infty}^{x}\frac{1}{\sqrt{2\pi}}e^{-y^{2}/2}\textrm{d}y$ denotes the standard cumulative Gaussian distribution function and $\phi(x)=\frac{1}{\sqrt{2\pi}}e^{-x^2/2}$ is the standard Gaussian density function. The detailed derivations of $\widetilde{\bm{I}}_{\widetilde{\bm{e}}_{u},b}$ follow those in \cite[Appendix~A]{crlbonebit} by accounting for the proposed double-sequence structure and (\ref{1bitqfunc}). Applying the chain rule, we can get the complex-valued Fisher information matrix as \cite{crlbonebit}
\begin{eqnarray}
\bm{I}_{\bm{e}_{u},b}&=&\frac{1}{4}\left(\left[\widetilde{\bm{I}}_{\widetilde{\bm{e}}_{u},b}\right]_{1:N,1:N}+\left[\widetilde{\bm{I}}_{\widetilde{\bm{e}}_{u},b}\right]_{(N+1):2N,(N+1):2N}\right)\nonumber\\
&+&\frac{\mathrm{j}}{4}\left(\left[\widetilde{\bm{I}}_{\widetilde{\bm{e}}_{u},b}\right]_{1:N,(N+1):2N}-\left[\widetilde{\bm{I}}_{\widetilde{\bm{e}}_{u},b}\right]_{(N+1):2N,1:N}\right).
\end{eqnarray}
For $n=0,\cdots,N-1$, we can then compute the $1$-bit CRLB as
\begin{equation}
\mathrm{var}\left(e^{\mathrm{j}\frac{2\pi\hat{\varepsilon}_u}{N}n}\left\vert\right. b\in\left\{1,\cdots,M_{\mathrm{tot}}\right\}\right)\geq\left[\bm{I}^{-1}_{\bm{e}_{u},b}\right]_{n,n}.
\end{equation}

In Fig.~\ref{fig:crlbfigure1ab}, we compare the derived CRLBs and the CFO estimation MSEs for our proposed auxiliary and sum-difference sequences based methods assuming $1$-bit ADCs. For simplicity, we assume single-path channels with a single UE. It is evident from Figs.~\ref{fig:crlbsubfigure1a} and \ref{fig:crlbsubfigure1b} that for $N=16$ and $N=32$, the CFO estimation MSEs of our proposed methods are close to the derived $1$-bit CRLBs especially under relatively high SNRs (e.g., $10$ dB). At high SNRs, the quantization distortion dominates the CFO estimation performance, resulting in error floors for both the derived CRLBs and the simulated MSEs.
\subsection{Performance analysis of auxiliary sequences under low-resolution ADCs}
Assuming that the input to the quantizer is IID Gaussian, we apply Bussgang's theorem and decompose the received synchronization signal samples in (\ref{ax0ori}) (i.e., synchronization time-slot $0$) into two uncorrelated components, given as
\begin{eqnarray}\label{ax0buss}
q^{0}_{u,b}[n]&=&(1-\kappa_{u})\left(e^{\mathrm{j}\frac{2\pi\varepsilon_u}{N}n}\sum_{\ell=0}^{L_u-1}\left[\bm{H}_u[\ell]\right]_{b,:}\bm{f}d_{0}[n-\ell]+w^{0}_{u,b}[n]\right)+v^{0}_{u,b}[n]\\
&=&(1-\kappa_{u})\left(e^{\mathrm{j}n(\mu_u-\theta+\delta)}\left[\textbf{\textsf{H}}_{u}\left(e^{-\mathrm{j}(\theta-\delta)}\right)\right]_{b,:}\bm{f}+w^{0}_{u,b}[n]\right)+v^{0}_{u,b}[n],
\end{eqnarray}
where $\bm{v}^{0}_{u,b}=\left[v^{0}_{u,b}[0],\cdots,v^{0}_{u,b}[N-1]\right]^{\mathrm{T}}$ represents the quantization noise vector with covariance matrix $\sigma^{2}_{u,\mathcal{Q}}\bm{I}_{N}$, and
\begin{equation}
\sigma^{2}_{u,\mathcal{Q}}=\kappa_{u}(1-\kappa_{u})\left(\mathbb{E}\left[\left|\left[\textbf{\textsf{H}}_u\left(e^{-\mathrm{j}(\theta-\delta)}\right)\right]_{b,:}\bm{f}\right|^{2}\right]+\sigma_u^2\right).
\end{equation}
Similarly, for auxiliary synchronization sequence $1$ (synchronization time-slot $1$), we express the corresponding quantized received synchronization signal samples as
\begin{equation}\label{ax1buss}
q^{1}_{u,b}[n]=(1-\kappa_{u})\left(e^{\mathrm{j}n(\mu_u-\theta-\delta)}\left[\textbf{\textsf{H}}_{u}\left(e^{-\mathrm{j}(\theta+\delta)}\right)\right]_{b,:}\bm{f}+w^{1}_{u,b}[n]\right)+v^{1}_{u,b}[n].
\end{equation}
The quantization noise vector $\bm{v}^{1}_{u,b}=\left[v^{1}_{u,b}[0],\cdots,v^{1}_{u,b}[N-1]\right]^{\mathrm{T}}$ has covariance matrix $\sigma^{2}_{u,\mathcal{Q}}\bm{I}_{N}$. Using the received signal samples from synchronization time-slots $0$ and $1$ and recalling that $\hat{b}$ is the selected receive antenna index at UE $u$,
\begin{align}
p^{0}_{u,\hat{b}}=&\left(\sum_{n=0}^{N-1}q^{0}_{u,\hat{b}}[n]\right)\left(\sum_{n=0}^{N-1}q^{0}_{u,\hat{b}}[n]\right)^{*}\\
=&\left[(1-\kappa_{u})\left(\left[\textbf{\textsf{H}}_{u}\left(e^{-\mathrm{j}(\theta-\delta)}\right)\right]_{\hat{b},:}\bm{f}\sum_{n=0}^{N-1}e^{\mathrm{j}n(\mu_u-\theta+\delta)}+\sum_{n=0}^{N-1}w^{0}_{u,\hat{b}}[n]\right)+\sum_{n=0}^{N-1}v^{0}_{u,\hat{b}}[n]\right]\nonumber\\
\times&\left[(1-\kappa_{u})\left(\left[\textbf{\textsf{H}}_{u}\left(e^{-\mathrm{j}(\theta-\delta)}\right)\right]_{\hat{b},:}\bm{f}\sum_{n=0}^{N-1}e^{\mathrm{j}n(\mu_u-\theta+\delta)}+\sum_{n=0}^{N-1}w^{0}_{u,\hat{b}}[n]\right)^{*}+\left(\sum_{n=0}^{N-1}v^{0}_{u,\hat{b}}[n]\right)^{*}\right]\label{axchhh0}\\
p^{1}_{u,\hat{b}}=&\left(\sum_{n=0}^{N-1}q^{1}_{u,\hat{b}}[n]\right)\left(\sum_{n=0}^{N-1}q^{1}_{u,\hat{b}}[n]\right)^{*}\\
=&\left[(1-\kappa_{u})\left(\left[\textbf{\textsf{H}}_{u}\left(e^{-\mathrm{j}(\theta+\delta)}\right)\right]_{\hat{b},:}\bm{f}\sum_{n=0}^{N-1}e^{\mathrm{j}n(\mu_u-\theta-\delta)}+\sum_{n=0}^{N-1}w^{1}_{u,\hat{b}}[n]\right)+\sum_{n=0}^{N-1}v^{1}_{u,\hat{b}}[n]\right]\nonumber\\
\times&\left[(1-\kappa_{u})\left(\left[\textbf{\textsf{H}}_{u}\left(e^{-\mathrm{j}(\theta+\delta)}\right)\right]_{\hat{b},:}\bm{f}\sum_{n=0}^{N-1}e^{\mathrm{j}n(\mu_u-\theta-\delta)}+\sum_{n=0}^{N-1}w^{1}_{u,\hat{b}}[n]\right)^{*}+\left(\sum_{n=0}^{N-1}v^{1}_{u,\hat{b}}[n]\right)^{*}\right].\label{axchhh1}
\end{align}
Using $p^{0}_{u,\hat{b}}$ in (\ref{axchhh0}) and $p^{1}_{u,\hat{b}}$ in (\ref{axchhh1}) to calculate $\alpha_u$ via (\ref{alpharatiometric}), the ratio measure is no longer a strict monotonic function of the CFO to be estimated due to the noise and quantization distortions. In this case, directly inverting the ratio metric may result in relatively large estimation errors. In the following, we first analytically characterize the impact of low-resolution ADCs on the proposed frequency synchronization methods. We then examine the robustness of our proposed methods to the quantization distortion in Section V-D.

By defining $c^{0}_{u,\hat{b}}=p^{0}_{u,\hat{b}}-p^{1}_{u,\hat{b}}$ as auxiliary channel $0$ and $c^{1}_{u,\hat{b}}=p^{0}_{u,\hat{b}}+p^{1}_{u,\hat{b}}$ as auxiliary channel $1$, we present the following lemma to characterize the variance of the CFO estimate under low-resolution quantization.
\begin{lemma1}
For the proposed auxiliary synchronization sequences based low-resolution frequency synchronization, the variance of the corresponding CFO estimate is
\begin{equation}\label{lllemma1}
\sigma^{2}_{\hat{\mu}_{u},\mathrm{ax}}=\left[\frac{\sin^{2}\left(\frac{N(\mu_u-\theta+\delta)}{2}\right)\left[\sin^{2}\left(\frac{\mu_u-\theta-\delta}{2}\right)-\sin^{2}\left(\frac{\mu_u-\theta+\delta}{2}\right)\right]}{N\left(1-\kappa_{u}\right)\left(\kappa_{u}+1/\gamma_{u,\mathrm{ax}}\right)}\right]^{-2}\frac{\left[1-\cos(\delta)\right]^{2}}{\sin^{2}(\delta)}\left(1+\alpha^{2}_{u}\right),
\end{equation}
\end{lemma1}
\emph{where}
\begin{equation}
\gamma_{u,\mathrm{ax}}=\mathbb{E}\left[\left|\left[\textbf{\textsf{H}}_u\left(e^{-\mathrm{j}(\theta\pm\delta)}\right)\right]_{\hat{b},:}\bm{f}\right|^{2}\right]\left/\right.\sigma_u^{2}.
\end{equation}
\begin{proof}
See Appendix \ref{proof1}.
\end{proof}As can be seen from (\ref{lllemma1}), the CFO estimation performance of the proposed method depends on various design parameters such as the sequence length $N$, the quantization NMSE $\kappa_{u}$, the reference frequency set $\left\{\theta,\delta\right\}$, and the received SNR $\gamma_{u,\mathrm{ax}}$.
\subsection{Performance analysis of sum-difference sequences under low-resolution ADCs}
By Bussgang's theorem, we decouple the received signal samples corresponding to the sum synchronization sequence in (\ref{sdseqdeng}) as
\begin{eqnarray}\label{sdseqdengbuss}
q^{\Sigma}_{u,b}[n]&=&\left(1-\kappa_{u}\right)\left(e^{\mathrm{j}\frac{2\pi\varepsilon_u}{N}n}\sum_{\ell=0}^{L_u-1}\left[\bm{H}_u[\ell]\right]_{b,:}\bm{f}d_{\Sigma}[n-\ell]+w^{\Sigma}_{u,b}[n]\right)+v^{\Sigma}_{u,b}[n]\\
&=&\left(1-\kappa_{u}\right)\left(e^{\mathrm{j}n(\mu_{u}-\eta)}\left[\textbf{\textsf{H}}_{u}\left(e^{-\mathrm{j}\eta}\right)\right]_{b,:}\bm{f}+w^{\Sigma}_{u,b}[n]\right)+v^{\Sigma}_{u,b}[n],
\end{eqnarray}
and the received signal samples for the difference synchronization sequence as
\begin{equation}\label{sdseqdengbussdelta}
q^{\Delta}_{u,b}[n]=\left\{\begin{array}{c}
                            \left(1-\kappa_{u}\right)\left(e^{\mathrm{j}n(\mu_{u}-\eta)}\left[\textbf{\textsf{H}}_{u}\left(e^{-\mathrm{j}\eta}\right)\right]_{b,:}\bm{f}+w^{\Delta}_{u,b}[n]\right)+v^{\Delta}_{u,b}[n],\hspace{2mm}n=0,1,\cdots,N/2-1 \\
                            \left(1-\kappa_{u}\right)\left(-e^{\mathrm{j}n(\mu_{u}-\eta)}\left[\textbf{\textsf{H}}_{u}\left(e^{-\mathrm{j}\eta}\right)\right]_{b,:}\bm{f}+w^{\Delta}_{u,b}[n]\right)+v^{\Delta}_{u,b}[n],\hspace{2mm}n=N/2,\cdots,N-1
                          \end{array}\right.
.
\end{equation}
The quantization noise vectors $\bm{v}_{u,b}^{\Sigma}=\left[v_{u,b}^{\Sigma}[0],\cdots,v_{u,b}^{\Sigma}[N-1]\right]^{\mathrm{T}}$ and $\bm{v}_{u,b}^{\Delta}=\big[v_{u,b}^{\Delta}[0],\cdots,v_{u,b}^{\Delta}[N-1]\big]^{\mathrm{T}}$ have the same covariance matrix $\sigma_{u,\mathcal{Q}}^{2}\bm{I}_{N}$, and
\begin{eqnarray}
\sigma_{u,\mathcal{Q}}^{2}=\kappa_u\left(1-\kappa_{u}\right)\left(\mathbb{E}\left[\left|\left[\textbf{\textsf{H}}\left(e^{-\mathrm{j}\eta}\right)\right]_{\hat{b},:}\bm{f}\right|^{2}\right]+\sigma_u^{2}\right).
\end{eqnarray}

For the selected $\hat{b}$-th receive antenna at UE $u$, we define $p^{\Sigma}_{u,\hat{b}}=\sum_{n=0}^{N-1}q^{\Sigma}_{u,\hat{b}}[n]$ and $p^{\Delta}_{u,\hat{b}} = \sum_{n=0}^{N-1}q^{\Delta}_{u,\hat{b}}[n]$ as the corresponding sum and difference channels. Based on (\ref{sumsyn0}) and (\ref{diffsyn0}), we obtain
\begin{align}
p^{\Sigma}_{u,\hat{b}} = & \left(1-\kappa_{u}\right)\left[\textbf{\textsf{H}}_{u}\left(e^{-\mathrm{j}\eta}\right)\right]_{\hat{b},:}\bm{f}\left(1+e^{\mathrm{j}(N/2)(\mu_{u}-\eta)}\right)e^{\mathrm{j}\frac{N/2-1}{2}(\mu_u-\eta)}\frac{\sin\left(\frac{N/2}{2}(\mu_{u}-\eta)\right)}{\sin\left((\mu_u-\eta)/2\right)}\nonumber\\
&+\left(1-\kappa_{u}\right)\sum_{n=0}^{N-1}w^{\Sigma}_{u,\hat{b}}[n]+\sum_{n=0}^{N-1}v^{\Sigma}_{u,\hat{b}}[n]\label{accr0}
\end{align}
\begin{align}
p^{\Delta}_{u,\hat{b}} = & \left(1-\kappa_{u}\right)\left[\textbf{\textsf{H}}_{u}\left(e^{-\mathrm{j}\eta}\right)\right]_{\hat{b},:}\bm{f}\left(1-e^{\mathrm{j}(N/2)(\mu_{u}-\eta)}\right)e^{\mathrm{j}\frac{N/2-1}{2}(\mu_u-\eta)}\frac{\sin\left(\frac{N/2}{2}(\mu_{u}-\eta)\right)}{\sin\left((\mu_u-\eta)/2\right)}\nonumber\\
&+\left(1-\kappa_{u}\right)\sum_{n=0}^{N-1}w^{\Delta}_{u,\hat{b}}[n]+\sum_{n=0}^{N-1}v^{\Delta}_{u,\hat{b}}[n].\label{accr1}
\end{align}
Using $p^{\Sigma}_{u,\hat{b}}$ in (\ref{accr0}) and $p^{\Delta}_{u,\hat{b}}$ in (\ref{accr1}) to compute $\beta_u$ via (\ref{betaratiometric}), the ratio metric is no longer a strict monotonic function of the CFO to be estimated. This observation is similar to that in Section IV-B for the proposed auxiliary sequences based approach. According to (\ref{accr0}) and (\ref{accr1}), we first provide the following lemma to reveal the variance of the CFO estimate assuming low-resolution quantization.
\begin{lemma2}
For the proposed sum-difference synchronization sequences based low-resolution frequency synchronization, the variance of the CFO estimate is given as
\begin{equation}\label{lemma2thereticalresult}
\sigma^{2}_{\hat{\mu}_{u},\mathrm{sd}}=\left[\frac{N\left(1-\kappa_{u}\right)\left|1+e^{\mathrm{j}(N/2)(\mu_{u}-\eta)}\right|^{2}\sin^{2}\left(\frac{N/2}{2}(\mu_{u}-\eta)\right)\csc^{4}\left(\frac{N}{4}\eta\right)}{16\left(\kappa_{u}+1/\gamma_{u,\mathrm{sd}}\right)\sin^{2}\left((\mu_u-\eta)/2\right)}\right]^{-1}\left(1+\beta^{2}_{u}\right),
\end{equation}
\end{lemma2}
\emph{where}
\begin{equation}
\gamma_{u,\mathrm{sd}}=\mathbb{E}\left[\left|\left[\textbf{\textsf{H}}_u\left(e^{-\mathrm{j}\eta}\right)\right]_{\hat{b},:}\bm{f}\right|^{2}\right]\left/\right.\sigma_u^{2}.
\end{equation}
The proof for Lemma 2 is similar to that for Lemma 1 by accounting for the sum and difference channels $p^{\Sigma}_{u,\hat{b}}$ and $p^{\Delta}_{u,\hat{b}}$. As can be seen from Lemma 2, the variance of the CFO estimate depends on the sequence length $N$, the quantization NMSE $\kappa_{u}$, the frequency difference $\triangle_u=\left|\mu_u-\eta\right|$, and the received SNR $\gamma_{u,\mathrm{sd}}$.
\subsection{Robustness to low-resolution quantization}
\begin{figure}
\centering
\subfigure[]{%
\includegraphics[width=.4\textwidth]{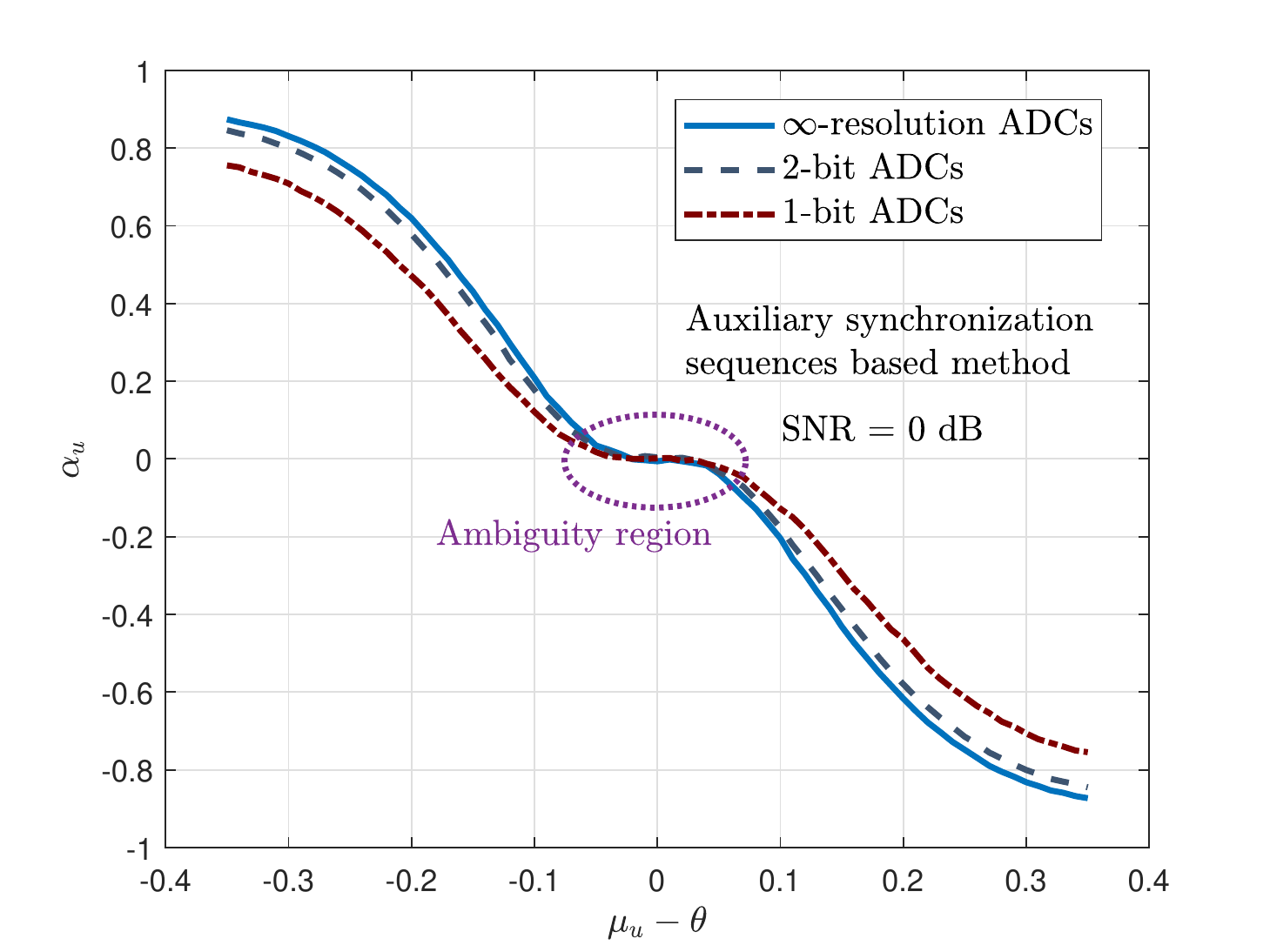}
\label{fig:impairedratio0}}
\subfigure[]{%
\includegraphics[width=.4\textwidth]{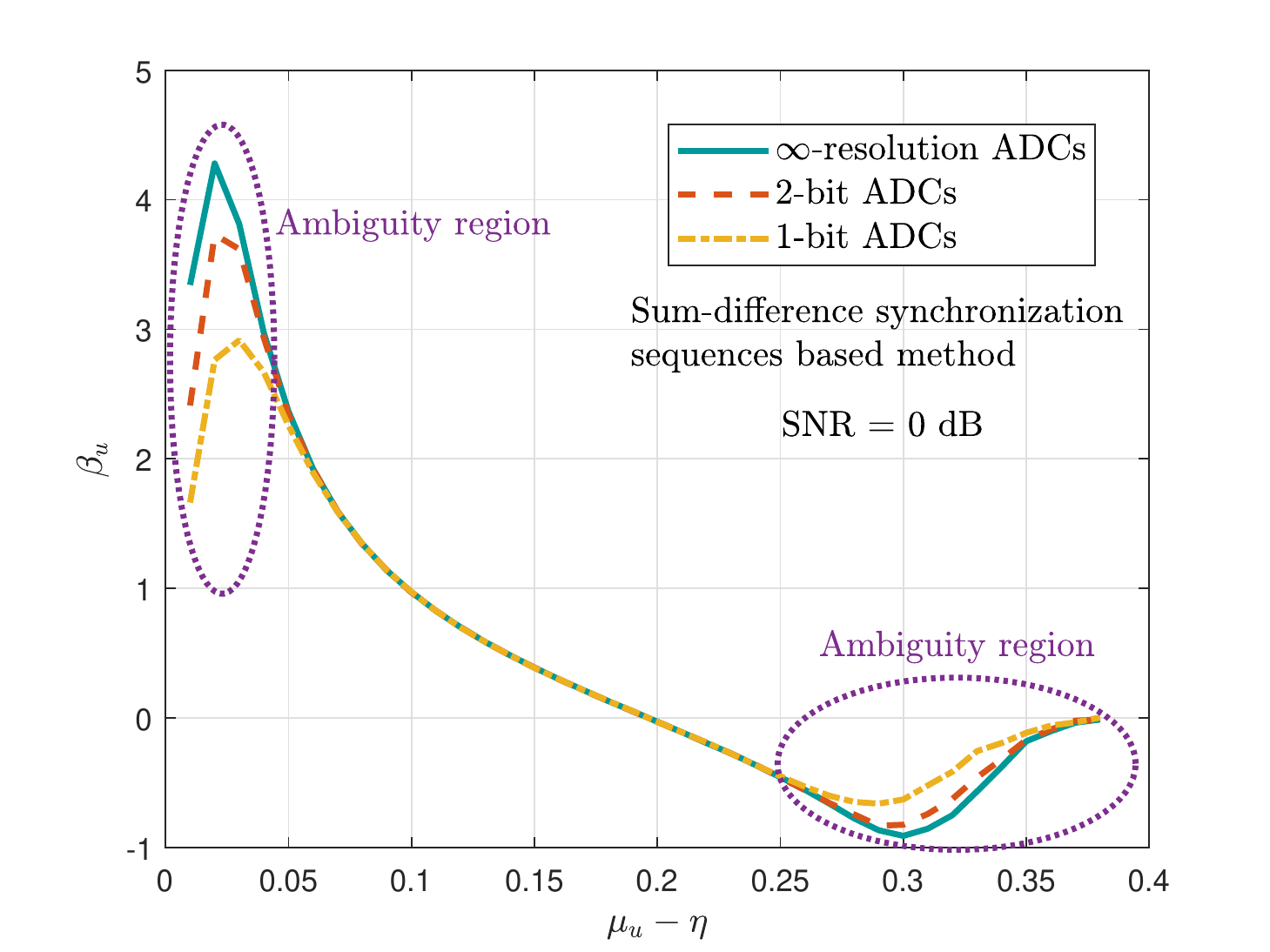}
\label{fig:impairedratio1}}
\caption{(a) Ratio metric versus $\mu_u-\theta$ for the proposed auxiliary sequences based double-sequence design; $0$ dB SNR is assumed with various ADC quantization resolutions. The ratio metric is no longer a monotonic function of the CFO to be estimated in the ambiguity region(s). (b) Ratio metric versus $\mu_u-\eta$ for the proposed sum-difference sequences based double-sequence design; $0$ dB SNR is assumed with various ADC quantization resolutions. The ratio metric is no longer a monotonic function of the CFO to be estimated in the ambiguity region(s).}
\label{fig:impairedratio}
\end{figure}
It is still difficult to analytically characterize the robustness of the proposed frequency synchronization methods to the quantization distortion, though their closed-form CFO estimation performances are presented in (\ref{lllemma1}) and (\ref{lemma2thereticalresult}). We therefore focus on empirical evaluations and plot the ratio metrics in Figs.~\ref{fig:impairedratio0} and \ref{fig:impairedratio1} for various quantization resolutions. As shown in Fig.~\ref{fig:impairedratio}, the monotonic properties do not hold for the ratio measures in the ambiguity regions such that directly inverting these ratio measures may result in large estimation errors. The ambiguity regions, however, are relatively small under various quantization resolutions. Further, except in the ambiguity regions, the actual ratio values under few-bit ADCs are close to those under infinite-resolution ADCs. Hence, by inverting the ratio measures obtained through few-bit ADCs, we can still expect promising CFO estimation performances for our proposed methods, which are numerically verified in Section VI.

As shown in (\ref{lllemma1}) and (\ref{lemma2thereticalresult}), the CFO estimation performances of our proposed methods also depend on certain double-sequence design parameter(s). For the auxiliary sequences based strategy, these parameters correspond to the reference frequency set $\left\{\theta,\delta\right\}$; for the sum-difference sequences based strategy, this parameter becomes $\eta$ (or the frequency difference $\triangle$). In addition to the sequence length $N$ and the received SNR, the double-sequence design parameter(s) can be further optimized to compensate for the quantization distortion. In Section V, we first use (\ref{lllemma1}) (or equivalently, (\ref{lemma2thereticalresult})) to formulate the optimization problems; we then conclude from the corresponding solutions that in contrast to conventional pilot/sequence-aided frequency synchronization methods, our proposed strategies are capable of exploiting more design degrees of freedom to further improve the low-resolution frequency synchronization performance.

\section{Practical Implementation of Low-Resolution Frequency Synchronization}
In this section, we explain the implementation procedure of our proposed double-sequence low-resolution frequency synchronization methods in practical cellular networks. We focus on a multi-user scenario with various quantization configurations. We exploit the analytical performance assessments from Lemmas 1 $\&$ 2 to configure the double-sequence design parameters such that for different deployment scenarios, the impact of low-resolution quantization distortion on the overall frequency synchronization performance can be minimized.
\subsection{Basic system setup and assumption}
In this paper, we consider a single cell serving multiple UEs. Note that our proposed designs can be extended to multiple cells by scrambling the synchronization channels with different physical cell identities (PCIs) to mitigate inter-cell interference. All active UEs are equipped with fully digital front ends and low-resolution ADCs. The ADC resolutions could be different for different UEs. The BS transmits a common synchronization signal to all active UEs via directional analog beamforming. In this paper, we focus on a given synchronization beam to explain and evaluate our proposed algorithms, though the BS forms multiple analog synchronization beams to ensure spatial coverage in practical systems. The design target here is to track all active UEs' CFOs without compromising much on each individual UE's CFO estimation performance. This can be accomplished by adjusting the frequency range of estimation of the common synchronization signal according to the distribution of all potential CFOs. In the following, we first formulate the multicast frequency synchronization problem as a min-max optimization problem assuming low-resolution quantization. By leveraging certain predefined system specific parameters, we transform the min-max problem into a minimization problem, and develop an adaptive algorithm to solve it. For the rest of this section, we employ the auxiliary sequences based method to illustrate the basic design procedure, which applies to the sum-difference sequences based design as well with moderate modifications.

\subsection{Optimization problem formulation of proposed low-resolution frequency synchronization}
To accommodate many UEs with satisfying low-resolution frequency synchronization performance, we need to incorporate the frequency range of estimation (e.g., $\left[\theta-\delta,\theta+\delta\right]$ in the auxiliary sequences based method and $\triangle$ in the sum-difference sequences based approach) into the optimization problem formulation. Define $\Theta$ as a codebook containing discrete values distributed within $[-1,1]$. Using (\ref{lllemma1}) and accounting for all $N_{\mathrm{UE}}$ active UEs, we can therefore formulate the corresponding low-resolution frequency synchronization problem as
\begin{eqnarray}\label{minmaxprob0}
\underset{\{\theta,\delta\}}{\mathrm{min}}\hspace{2mm}\underset{\forall u}{\mathrm{max}}&&\left\{\sigma^{2}_{\hat{\mu}_u,\mathrm{ax}}\right\}\\
\textrm{subject to}&&\theta\in\Theta\nonumber\\
&&\delta=2k'\pi/N, k'=1,\cdots,\frac{N}{4}\nonumber\\
&&u\in\{1,\cdots,N_{\mathrm{UE}}\},\nonumber
\end{eqnarray}
which is a min-max optimization problem and difficult to solve. To simplify (\ref{minmaxprob0}), we first define $\kappa_{\mathrm{ax}}$, $\mu_{\mathrm{ax}}$, $\alpha_{\mathrm{ax}}$ and $\gamma_{\mathrm{ax}}$ as system specific parameters. They are predefined and can be flexibly configured by the network controller according to measurements and system statistics. We use $\kappa_{\mathrm{ax}}$, $\mu_{\mathrm{ax}}$, $\alpha_{\mathrm{ax}}$ and $\gamma_{\mathrm{ax}}$ to replace $\kappa_{u}$, $\mu_{u}$, $\alpha_{u}$ and $\gamma_{u,\mathrm{ax}}$ in (\ref{lllemma1}) and obtain
\begin{equation}\label{minmaxrwt}
\sigma^{2}_{\hat{\mu}_{\mathrm{ax}}}=\left[\frac{\sin^{2}\left(\frac{N(\mu_{\mathrm{ax}}-\theta+\delta)}{2}\right)\left[\sin^{2}\left(\frac{\mu_{\mathrm{ax}}-\theta-\delta}{2}\right)-\sin^{2}\left(\frac{\mu_{\mathrm{ax}}-\theta+\delta}{2}\right)\right]}{N\left(1-\kappa_{\mathrm{ax}}\right)\left(\kappa_{\mathrm{ax}}+1/\gamma_{\mathrm{ax}}\right)}\right]^{-2}\frac{\left[1-\cos(\delta)\right]^{2}}{\sin^{2}(\delta)}\left(1+\alpha^{2}_{\mathrm{ax}}\right).
\end{equation}

We can then rewrite the optimization problem as
\begin{eqnarray}\label{minmaxrft}
\underset{\{\theta,\delta\}}{\mathrm{minimize}} &&\left\{\sigma^{2}_{\hat{\mu}_{\mathrm{ax}}}\right\}\\
\textrm{subject to} &&\theta\in\Theta\nonumber\\
&&\delta=2k'\pi/N, k'=1,\cdots,\frac{N}{4}.\nonumber
\end{eqnarray}
We formulate (\ref{minmaxrft}) by treating all active UEs as a single virtual UE with common system specific parameters $\kappa_{\mathrm{ax}}$, $\mu_{\mathrm{ax}}$, $\alpha_{\mathrm{ax}}$, $\gamma_{\mathrm{ax}}$, and a given analog synchronization beam $\bm{f}$. We need to carefully select $\kappa_{\mathrm{ax}}$, $\mu_{\mathrm{ax}}$, $\alpha_{\mathrm{ax}}$ and $\gamma_{\mathrm{ax}}$ such that $\sigma^{2}_{\hat{\mu}_{\mathrm{ax}}}$ for the virtual UE can represent each individual UE's error variance, i.e., $\sigma^{2}_{\hat{\mu}_u,\mathrm{ax}}$ ($u\in\left\{1,\cdots,N_{\mathrm{UE}}\right\}$), as much as possible. We present one example of how to determine the system specific parameters in Section V-C. Note that if $\kappa_{u}$, $\mu_{u}$, $\alpha_{u}$ and $\gamma_{u,\mathrm{ax}}$ ($u\in\left\{1,\cdots,N_{\mathrm{UE}}\right\}$) vary significantly from UE to UE, the problem formulation in (\ref{minmaxrwt}) and (\ref{minmaxrft}) may not be accurate. This in turn, may result in poor frequency synchronization performance in the multi-user scenario. In Section VI, we use several numerical examples to characterize this aspect. Finally, we solve (\ref{minmaxrft}) by optimizing the double-sequence parameters such that the estimation error variance of the virtual UE is minimized.
\subsection{Design procedure of proposed low-resolution frequency synchronization}
According to (\ref{minmaxrwt}), solving (\ref{minmaxrft}) requires the BS to have explicit knowledge of $\kappa_{\mathrm{ax}}$, $\mu_{\mathrm{ax}}$ ($\alpha_{\mathrm{ax}}$) and $\gamma_{\mathrm{ax}}$. Though different UEs may have different ADC resolutions, we use a single $\kappa_{\mathrm{ax}}$ to characterize all active UEs. In Section VI, we set $\kappa_{\mathrm{ax}}=0.1175$ (i.e., $2$-bit ADCs) when simulating the multi-user scenario. Further, $\gamma_{\mathrm{ax}}$ can be obtained by averaging over the received signal-to-interference-plus-noise ratios (SINRs) of all UEs. Selecting a good $\mu_{\mathrm{ax}}$ is also important for the BS to determine a proper frequency range of estimation that can capture as many UEs' CFOs as possible. In this paper, we assume that the BS or the network controller configures $\mu_{\mathrm{ax}}$ using long-term system statistics such as those from previously connected UEs.

\begin{algorithm}
  \caption{Auxiliary sequences based low-resolution frequency synchronization design}
  \begin{algorithmic}\\
\State \textbf{BS-SIDE PROCESSING}
\begin{itemize}
  \item[1]:~Configuring $\kappa_{\mathrm{ax}}$, $\mu_{\mathrm{ax}}$, $\alpha_{\mathrm{ax}}$ and $\gamma_{\mathrm{ax}}$ based on long-term measurements and system statistics.
  \item[2]:~Computing $\sigma^{2}_{\hat{\mu}_{\mathrm{ax}}}$ according to (\ref{minmaxrwt}) and finding $\theta_{\mathrm{opt}}$ and $\delta_{\mathrm{opt}}$ that minimizes $\sigma^{2}_{\hat{\mu}_{\mathrm{ax}}}$ following (\ref{minmaxrft}).
  \item[3]:~If necessary, conveying the selected $\theta_{\mathrm{opt}}$ and $\delta_{\mathrm{opt}}$ to UEs after their timing synchronization.
  \item[4]:~Constructing $\bm{d}_{0}$ and $\bm{d}_{1}$ according to (\ref{axvia0}) and (\ref{axvia1}), and maps their frequency-domain counterparts on given subcarriers.
  \item[5]:~Probing the synchronization signal at given synchronization time-slots.
\end{itemize}
\State \textbf{UE-SIDE PROCESSING (UE $u$, $u\in\{1,\cdots,N_{\mathrm{UE}}\}$)}
\begin{itemize}
\item[i]:~Preprocessing: timing synchronization and discarding the CP.
\item[ii]:~Computing the received synchronization signal strengths $p^{0}_{u,\hat{b}}$, $p^{1}_{u,\hat{b}}$ and the ratio measure $\alpha_u$ according to (\ref{alpharatiometric}).
\item[iii]:~Inverting the ratio measure $\alpha_u$ according to (\ref{howtoinvertr}) to obtain the CFO estimate $\hat{\mu}_{u}$.
\item[iv]:~Compensating the received signal samples with the estimated CFO.
\end{itemize}
  \end{algorithmic}
\end{algorithm}
We summarize the detailed design procedure in Algorithm 1. As can be seen from Algorithm 1, the BS needs to convey the optimized double-sequence design parameters (e.g., $\theta_{\mathrm{opt}}$ and $\delta_{\mathrm{opt}}$ in the auxiliary sequences based method) to the UE so that the UE can execute the ratio metric inversion. This requires additional signaling support from the BS to the UE, which can still be implemented in the initial access process after the UE completes the symbol/frame timing synchronization and PCI detection. To reduce this signaling overhead, the BS can optimize the double-sequence design parameters in a semi-static manner, using long-term measurements and system statistics.
\section{Numerical Results}
In this section, we evaluate our proposed auxiliary and sum-difference sequences based frequency synchronization methods assuming low-resolution (e.g., $1$-$4$ bits) ADCs. Both the BS and UE employ a uniform linear array (ULA) with inter-element spacing $\lambda/2$, where $\lambda$ represents the wavelength corresponding to the carrier frequency. Throughout the simulation section, we assume that $N_{\mathrm{tot}}=16$ and $M_{\mathrm{tot}}=8$. The BS covers $120^{\circ}$ angular range $[-60^{\circ},60^{\circ}]$ around boresight ($0^{\circ}$), and the UE monitors the entire $180^{\circ}$ angular region $[-90^{\circ},90^{\circ}]$ around boresight ($0^{\circ}$). We consider both narrowband and wideband mmWave channels to evaluate our proposed algorithms, incorporating a single UE or multiple UEs.
\subsection{Narrowband mmWave channels with a single UE}
\begin{figure}
\begin{center}
\subfigure[]{%
\includegraphics[width=.4\textwidth]{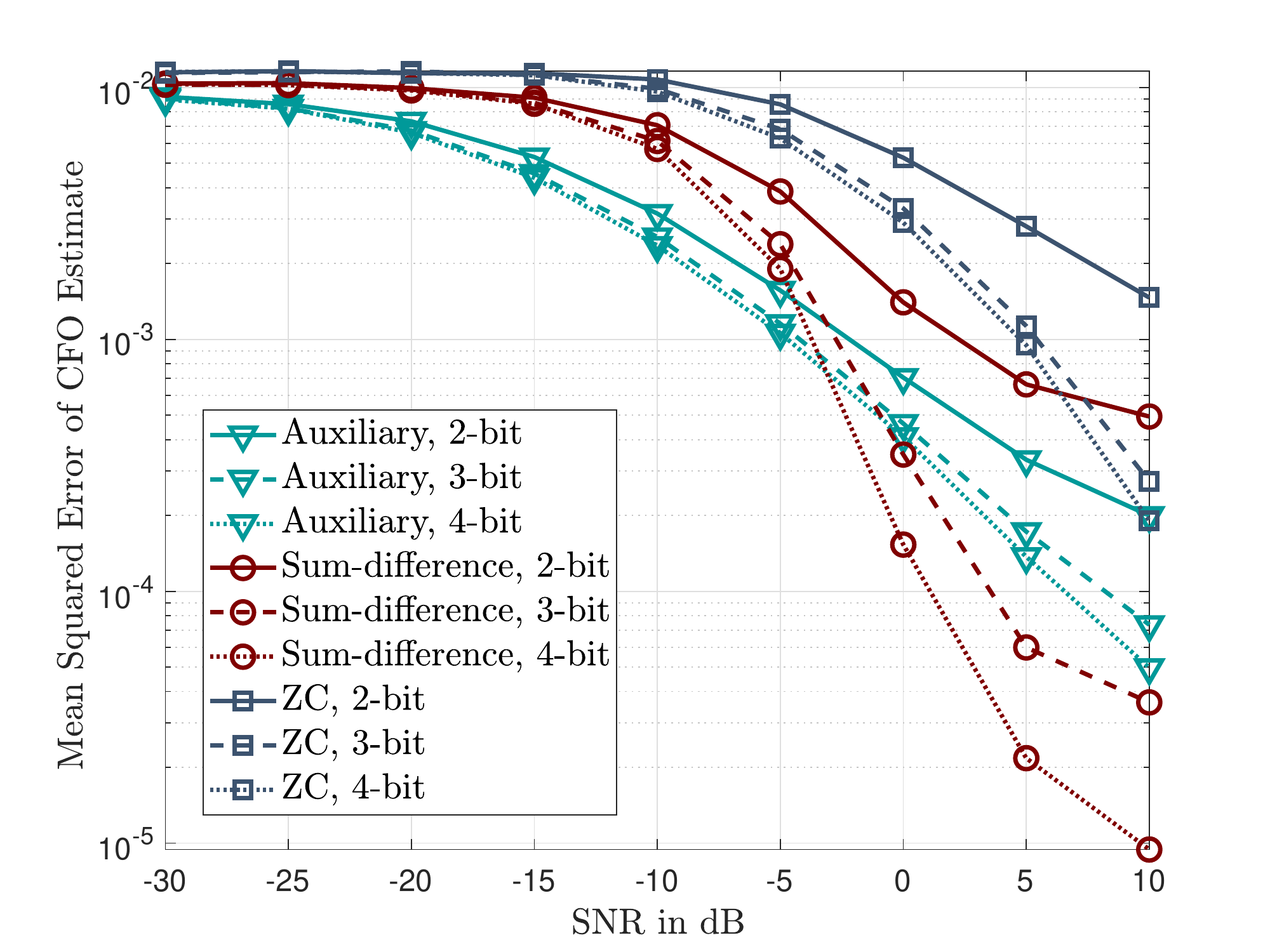}
\label{fig:simsubfigure0a}}
\hspace{-3.5mm}
\subfigure[]{%
\includegraphics[width=.4\textwidth]{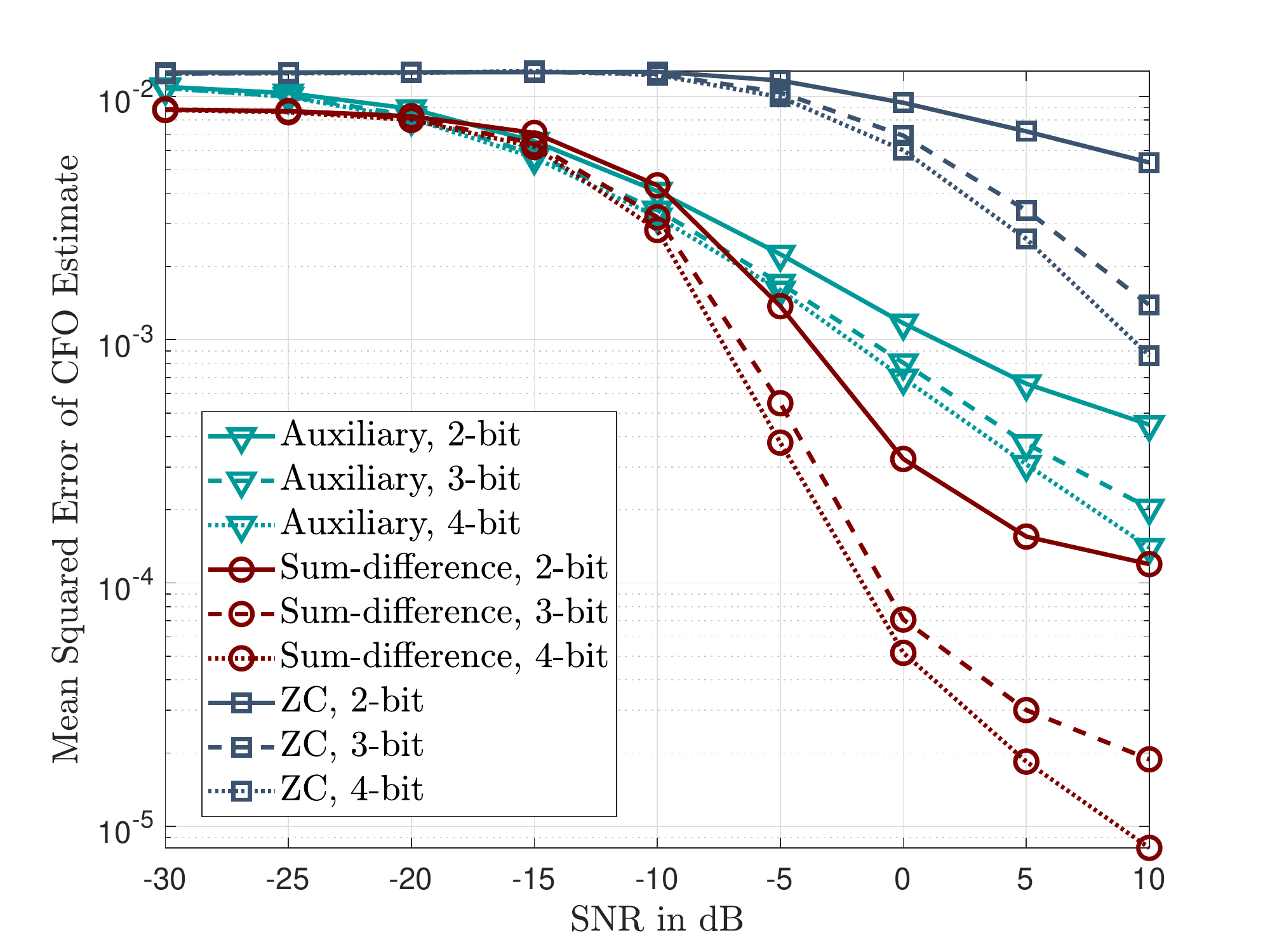}
\label{fig:simsubfigure0b}}
\caption{Mean squared errors of the CFO estimates obtained via the ZC sequence based method and our proposed auxiliary and sum-difference sequences based methods are plotted against various SNRs. Narrowband Rician channels are assumed with $13.2$~dB Rician $K$-factor and a single UE. The assumed quantization resolutions for the ADCs are $2$, $3$ and $4$ bits. (a) $0.6$ normalized CFO. (b) $0.95$ normalized CFO.}
\label{fig:simfigure0ab}
\end{center}
\end{figure}

In Fig.~\ref{fig:simfigure0ab}, we evaluate our proposed strategies in narrowband channels with a single UE. Specifically, we employ the Rician channel model, which is expressed as
\begin{eqnarray}
\bm{H} = \sqrt{\frac{K}{1+K}}\bm{H}_{\mathrm{LOS}}+\sqrt{\frac{\mathsf{1}}{1+K}}\bm{H}_{\mathrm{NLOS}},
\end{eqnarray}
where $\bm{H}_{\mathrm{LOS}}$ and $\bm{H}_{\mathrm{NLOS}}$ represent line-of-sight (LOS) and non-LOS (NLOS) channel components. We set the number of NLOS channel components as $5$ and the Rician $K$-factor value as $13.2$ dB. From the observation in \cite{rician}, the $13.2$ dB Rician $K$-factor characterizes the mmWave channels' conditions the best in an urban wireless channel topography. We also simulate the conventional ZC sequence based CFO estimation approach for comparison. We set the root index as $24$ for the employed ZC sequence.

We plot the MSEs of the CFO estimates in Figs.~\ref{fig:simsubfigure0a} and \ref{fig:simsubfigure0b}, in which we set the normalized CFOs as $0.6$ and $0.95$. It can be observed from Figs.~\ref{fig:simsubfigure0a} and \ref{fig:simsubfigure0b} that under various quantization resolutions, our proposed auxiliary sequences and sum-difference sequences based methods exhibit superior CFO estimation performances over the ZC sequence based design. For $0.95$ normalized CFO in Fig.~\ref{fig:simsubfigure0b}, the sum-difference sequences based approach shows better performance than the auxiliary sequences based method under various SNRs.

\subsection{Wideband mmWave channels with a single UE}
\begin{figure}
\centering
\includegraphics[width=5.05in]{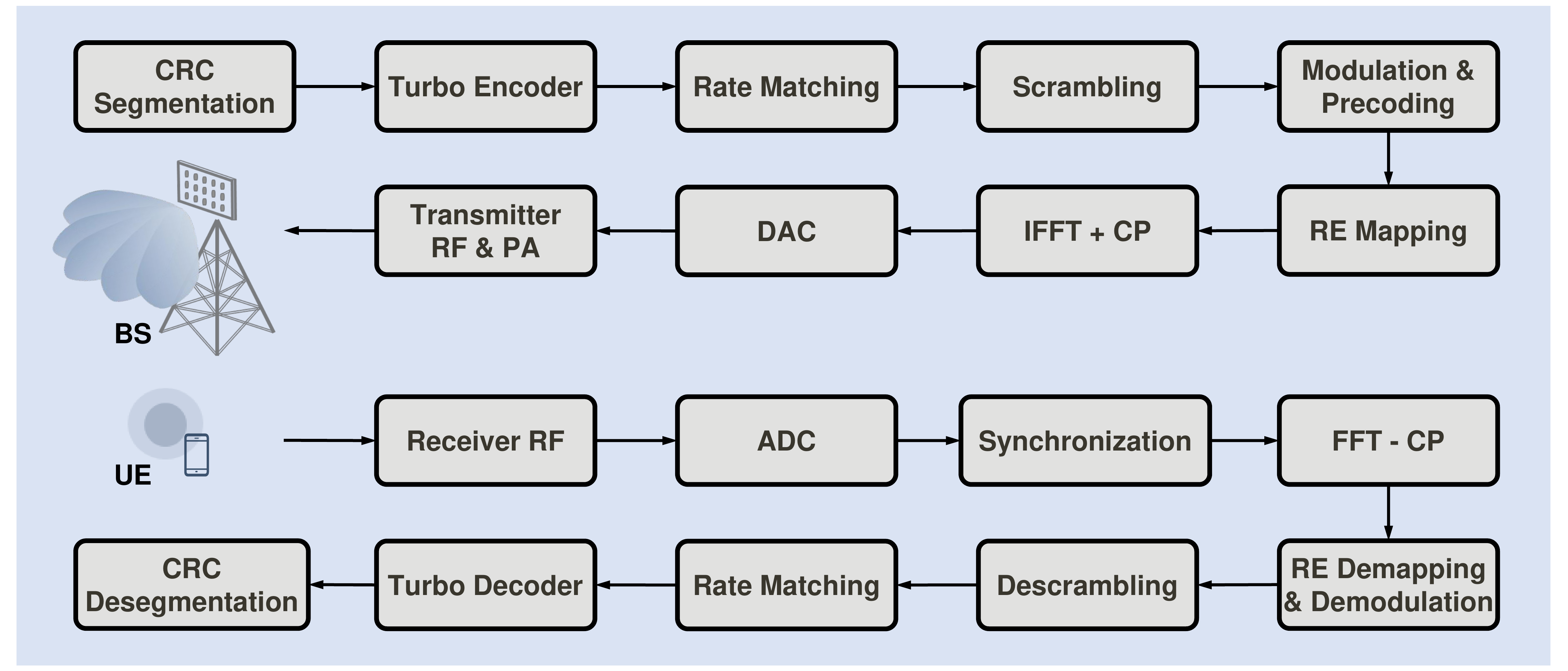}
\caption{Block diagram of employed link-level simulator for evaluating our proposed low-resolution frequency synchronization methods in wideband mmWave channels.}
\label{fig:figureLLS}
\end{figure}
\begin{table*}[t] \centering
\caption{Simulation assumptions and parameters.}\label{tracking_comp}
\begin{tabular}{||c||c||}
  \hline
  \textbf{SYSTEM PARAMETERS} & \textbf{SIMULATION ASSUMPTIONS} \\
  \hline
  Carrier frequency & $30$ GHz \\
  \hline
  System bandwidth & $80$ MHz \\
  \hline
  FFT size & $1024$ \\
  \hline
  Subcarrier spacing & $120$ KHz \\
  \hline
  OFDM symbol duration ($\mu$s) & $8.33$ \\
  \hline
  CP length ($\mu$s) & 0.82 \\
  \hline
  Channel coding & Turbo \\
  \hline
  MCS & QPSK, code rate $1/2$ \\
  \hline
  Channel model & 3GPP 5G NR TDL-A \cite{5gnrtdla} \\
  \hline
  Channel estimation & Ideal \\
  \hline
\end{tabular}
\end{table*}
\begin{figure}
\begin{center}
\subfigure[]{%
\includegraphics[width=.4\textwidth]{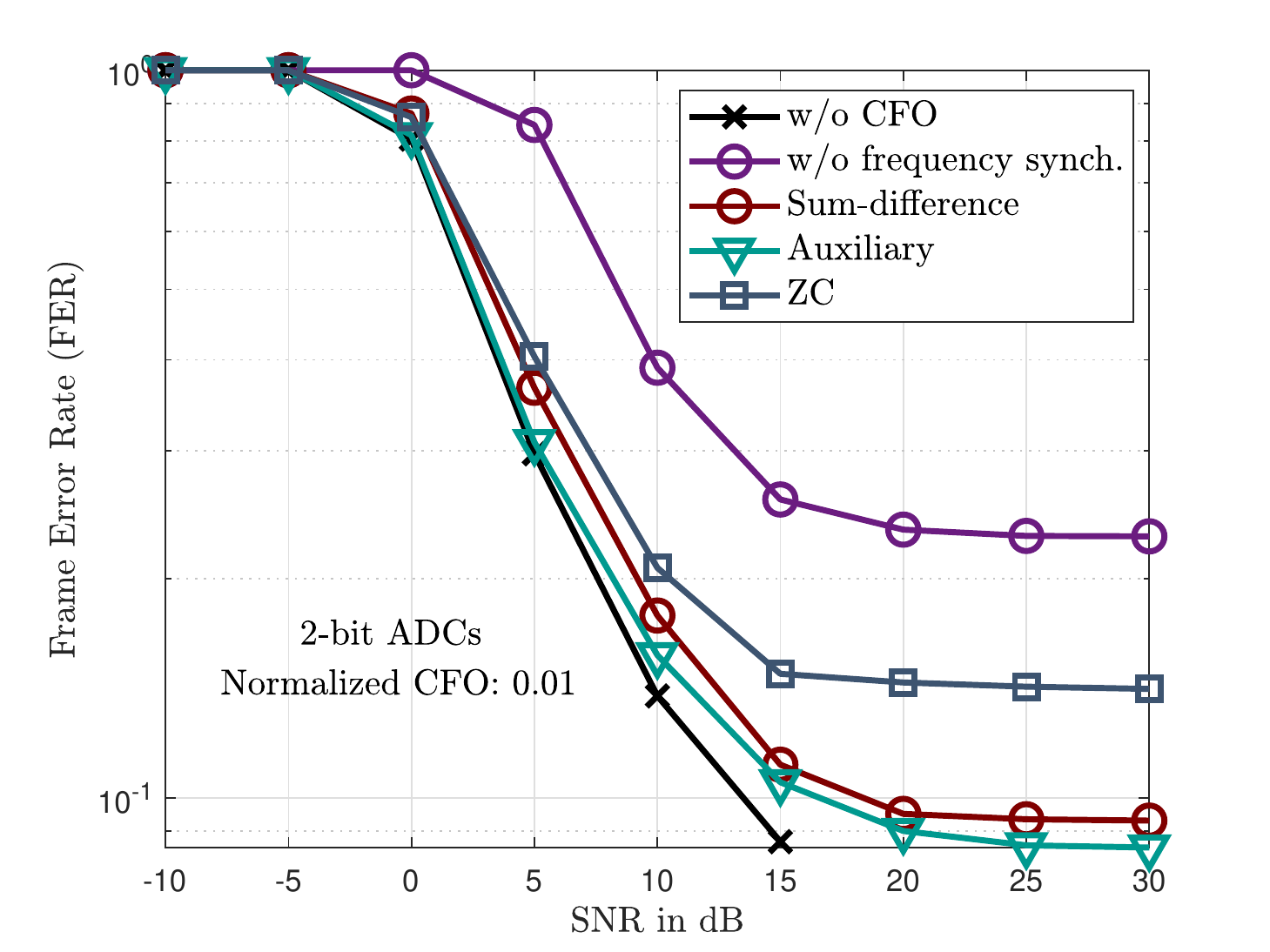}
\label{fig:simsubfigureFERa}}
\hspace{-3.5mm}
\subfigure[]{%
\includegraphics[width=.4\textwidth]{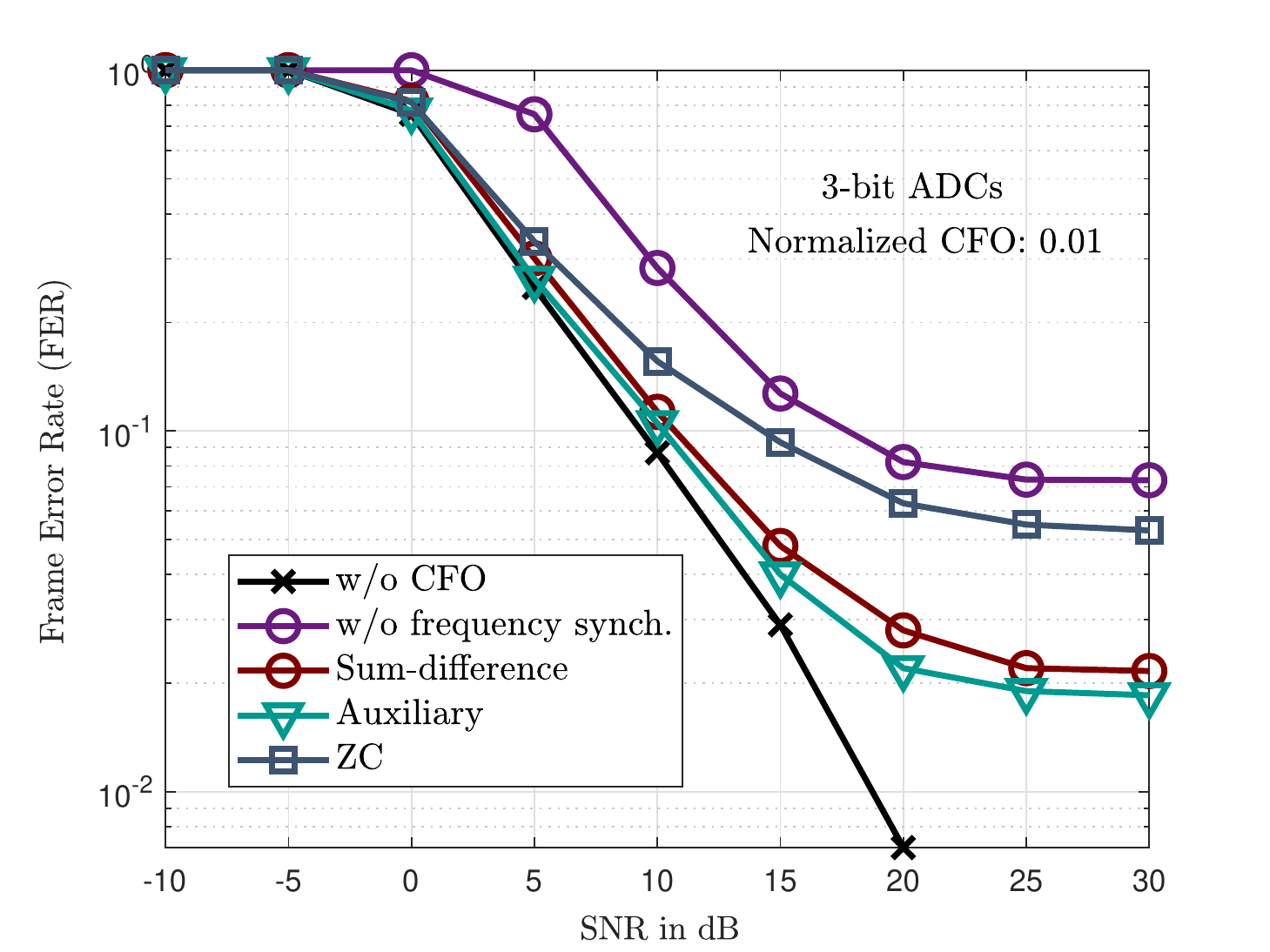}
\label{fig:simsubfigureFERb}}
\caption{Frame error rate performances of our proposed auxiliary and sum-difference sequences based low-resolution frequency synchronization methods. The ZC sequence based method is evaluated for comparison. The perfect case without CFO and the case with CFO but without any frequency synchronization are plotted as benchmarks. The normalized CFO is set to $0.01$. Other simulation assumptions are given in Table I. (a) $2$-bit ADCs. (b) $3$-bit ADCs.}
\label{fig:simfigureFERab}
\end{center}
\end{figure}
Next, we evaluate the frame error rate (FER) of our proposed methods assuming few-bit ADCs. We present the block diagram of our link-level simulator in Fig.~\ref{fig:figureLLS}, which includes a complete chain of Turbo coding/decoding, OFDM and DAC/ADC modules. We assume perfect channel estimation, frame timing synchronization and infinite-resolution DACs. We adopt the 3GPP 5G tapped delay line (TDL-A) wideband channel model \cite{5gnrtdla} into the link-level simulation assuming $30$ GHz carrier frequency and a $80$ MHz RF bandwidth. We provide other simulation assumptions such as the assumed OFDM numerology and the modulation and coding scheme (MCS) in Table I. According to these settings, a $0.1$ normalized CFO corresponds to a $1.95$~MHz time-domain CFO or $65$ ppm of the $30$ GHz carrier frequency. In Figs.~\ref{fig:simsubfigureFERa} and \ref{fig:simsubfigureFERb}, we plot the FER performances assuming both $2$-bit and $3$-bit ADCs. As can be seen from Fig.~\ref{fig:simfigureFERab}, our proposed auxiliary and sum-difference sequences based frequency synchronization strategies significantly outperform the ZC sequence based CFO estimation method. Further, in the low-to-medium SNR regime (e.g., $-10$ dB $\sim10$ dB), our proposed methods show error rate performances close to the ideal case without CFO. For our proposed methods and the ZC sequence based strategy, we can observe error floors at high SNRs.

\begin{figure}
\begin{center}
\subfigure[]{%
\includegraphics[width=.4\textwidth]{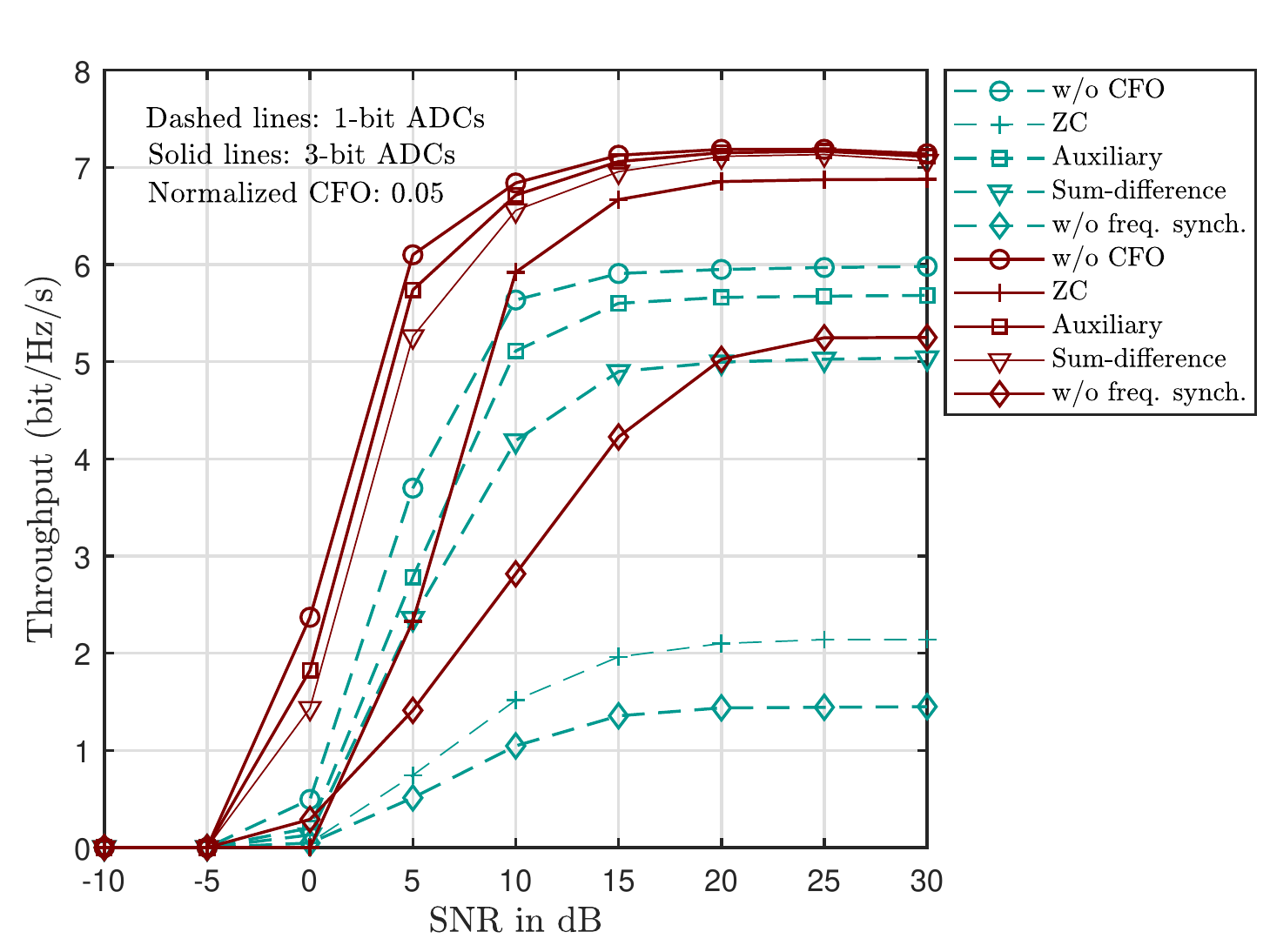}
\label{fig:simsubfigureThputa}}
\hspace{-3.5mm}
\subfigure[]{%
\includegraphics[width=.4\textwidth]{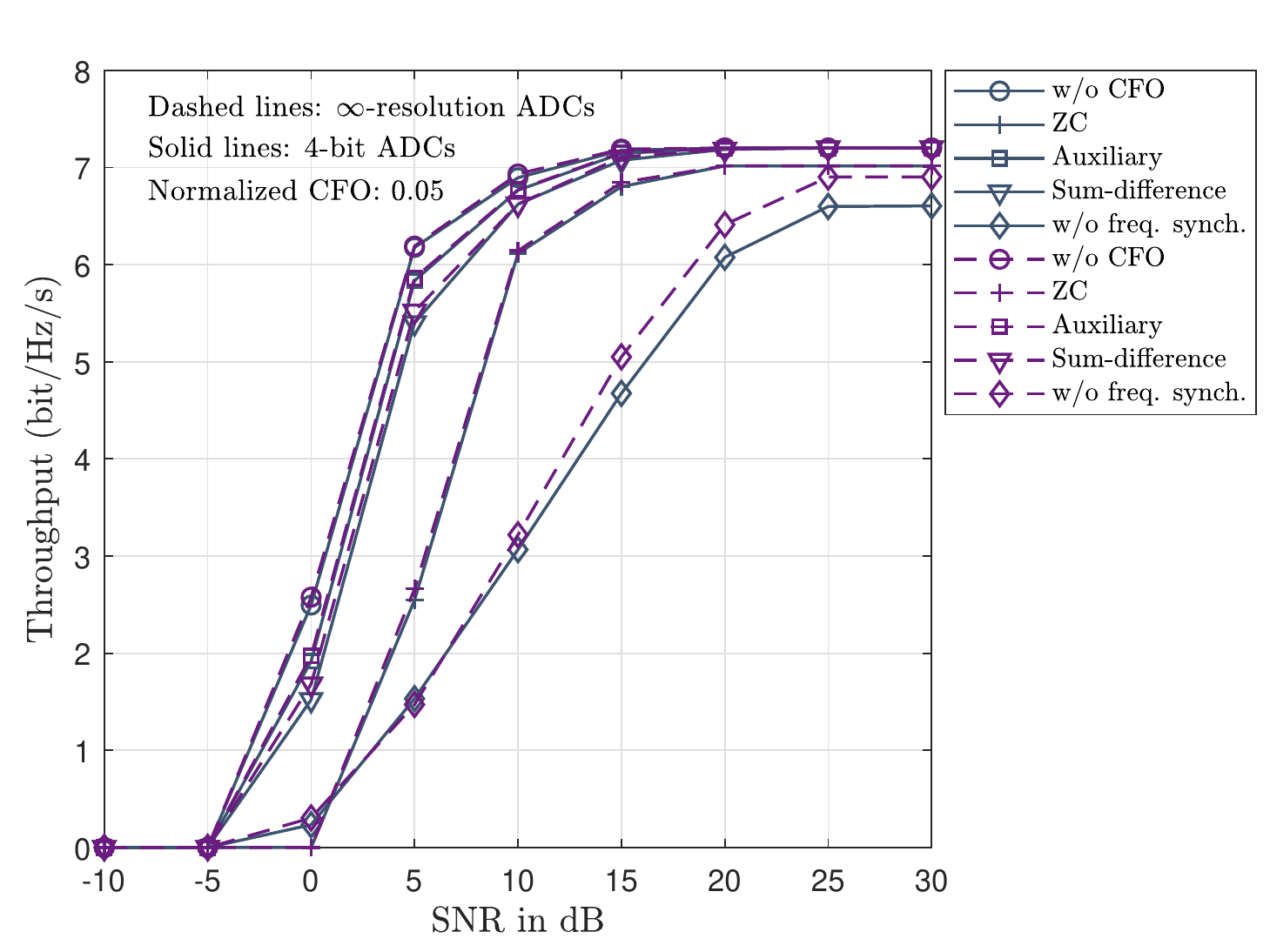}
\label{fig:simsubfigureThputb}}
\caption{Throughput performances versus SNRs of our proposed auxiliary and sum-difference sequences based low-resolution frequency synchronization methods. The ZC sequence based method is evaluated for comparison. The perfect case without CFO and the case with CFO but without any frequency synchronization are plotted as benchmarks. The normalized CFO is set to $0.05$. Other simulation assumptions are given in Table I. (a) $1$-bit and $3$-bit ADCs. (b) $4$-bit and infinite-resolution ADCs.}
\label{fig:simfigureThputab}
\end{center}
\end{figure}
In Fig.~\ref{fig:simfigureThputab}, we evaluate the throughput performances for our proposed methods assuming various quantization resolutions. We compare $1$-bit and $3$-bit ADCs in Fig.~\ref{fig:simsubfigureThputa} and observe that our proposed methods outperform the ZC sequence based strategy (similar to Fig.~\ref{fig:simfigureFERab}) for $0.05$ normalized CFO. For instance, at $20$ dB SNR, the throughput of our proposed sum-difference sequences based method is $3$ bit/s/Hz higher than that of the ZC sequence based strategy assuming $1$-bit ADCs. The corresponding throughput gain is $150\%$. With increase in the quantization resolution, this performance gap reduces. In Fig.~\ref{fig:simsubfigureThputb}, we examine our proposed approaches under $4$-bit and infinite-resolution ADCs. The performance differences between $4$-bit quantization and infinite-resolution quantization are marginal.

\begin{figure}
\begin{center}
\subfigure[]{%
\includegraphics[width=.4\textwidth]{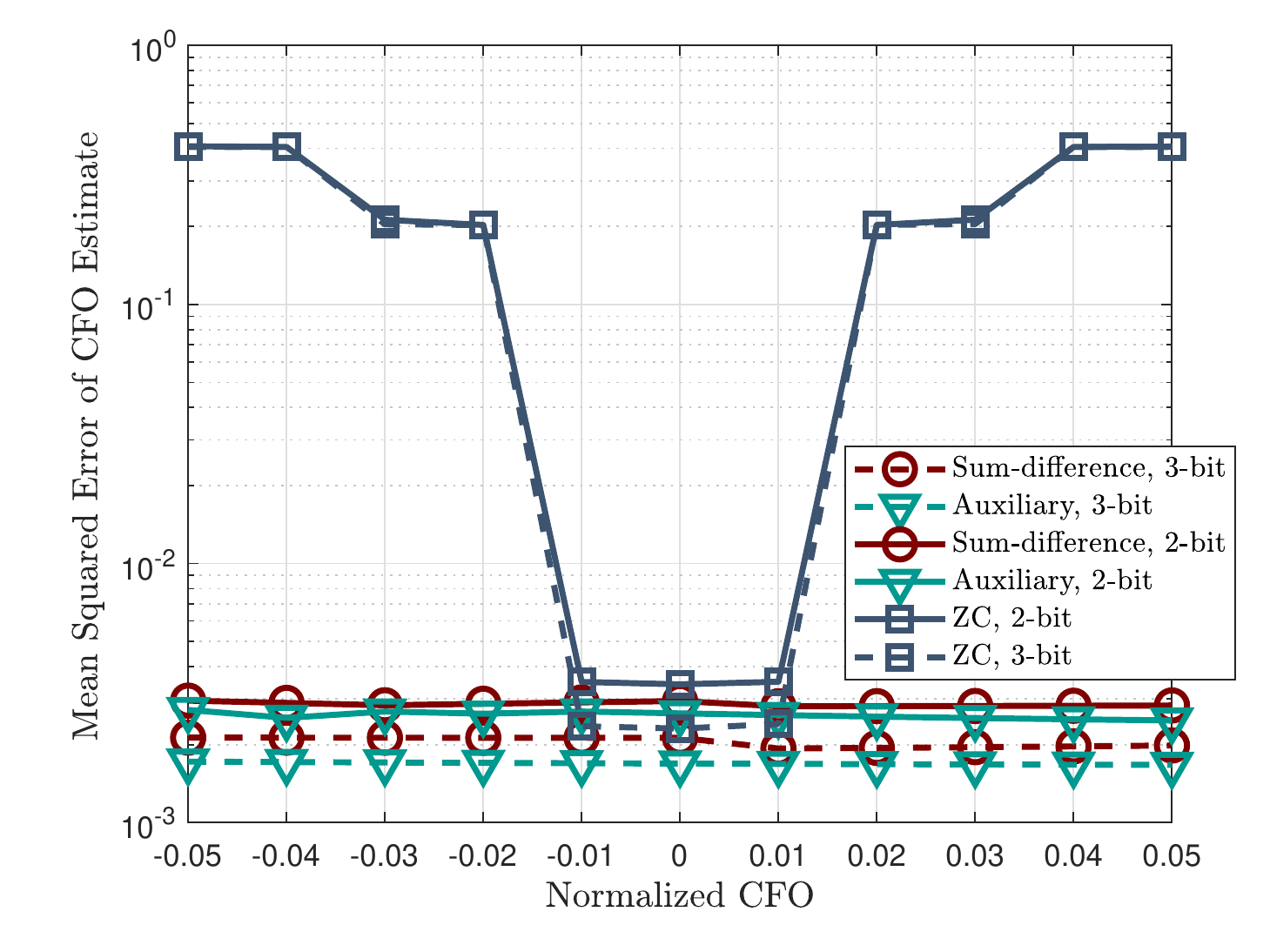}
\label{fig:simsubfigureMSECFOa}}
\hspace{-3.5mm}
\subfigure[]{%
\includegraphics[width=.4\textwidth]{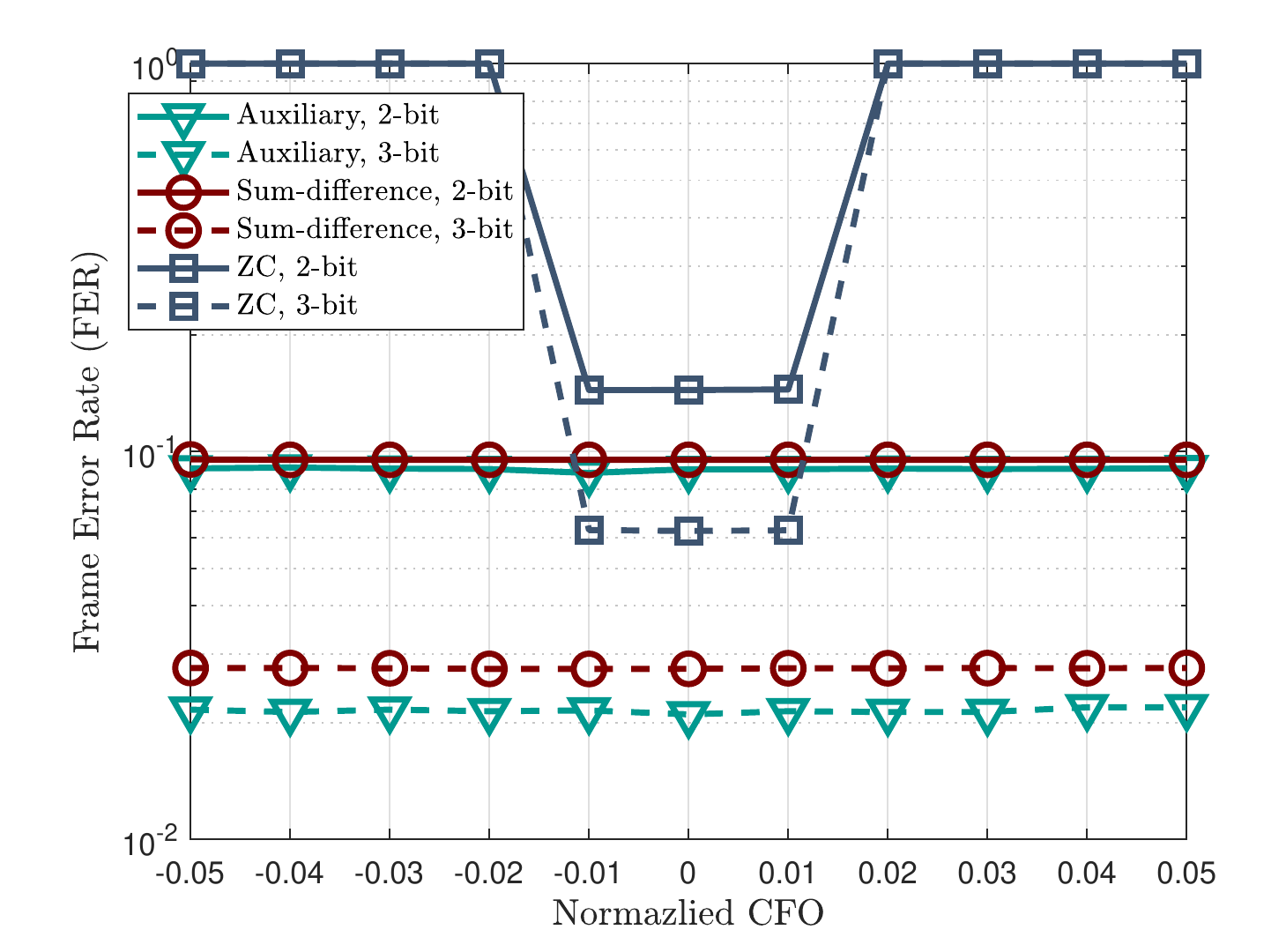}
\label{fig:simsubfigureMSECFOb}}
\caption{Low-resolution frequency synchronization performance evaluations under various normalized CFO values. The assumed quantization resolutions for the ADCs are $2$ and $3$ bits with $20$ dB SNR. (a) Mean squared errors of the CFO estimates obtained via the ZC sequence based method and our proposed auxiliary and sum-difference sequences based methods. (b) Frame error rate performances of the ZC sequence based method and our proposed auxiliary and sum-difference sequences based methods.}
\label{fig:simfigureMSECFOab}
\end{center}
\end{figure}
In Fig.~\ref{fig:simsubfigureMSECFOa}, we plot the MSEs of the CFO estimates for various frequency synchronization methods. It is evident from Fig.~\ref{fig:simsubfigureMSECFOa} that our proposed CFO estimation strategies are better than the conventional ZC sequence based approach for a wide range of frequencies. Further, the MSEs of our proposed methods are almost identical across the $[-0.05,0.05]$ frequency range of interest. This is because our proposed methods can provide super-resolution CFO estimates as long as the monotonic properties hold. Similar observations are obtained in Fig.~\ref{fig:simsubfigureMSECFOb}, in which the FER performances are evaluated against various normalized CFOs.
\subsection{Wideband mmWave channels with multiple UEs}
\begin{figure}
\begin{center}
\subfigure[]{%
\includegraphics[width=.4\textwidth]{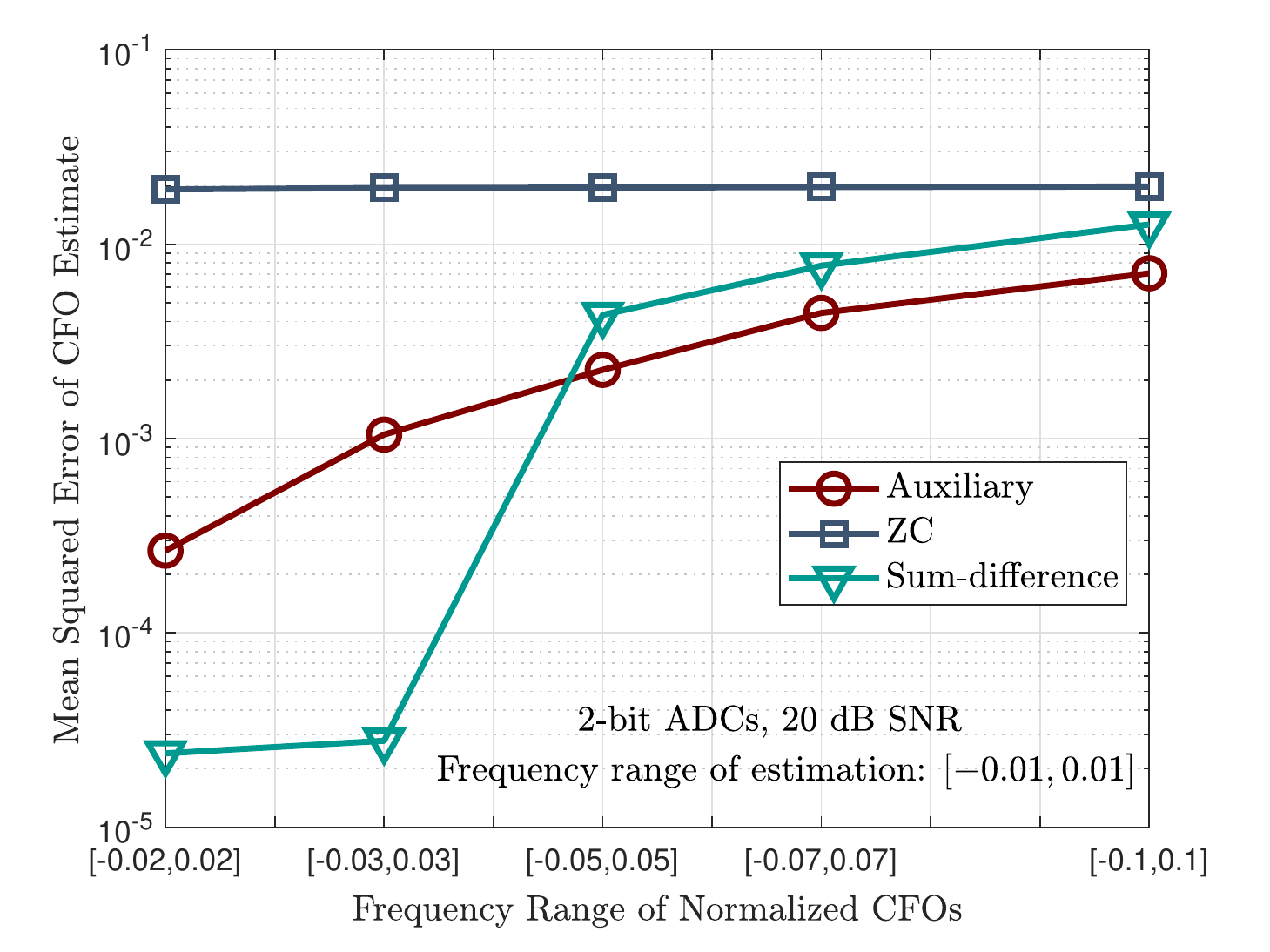}
\label{fig:simsubfigureMSECFOmutwobita}}
\hspace{-3.5mm}
\subfigure[]{%
\includegraphics[width=.4\textwidth]{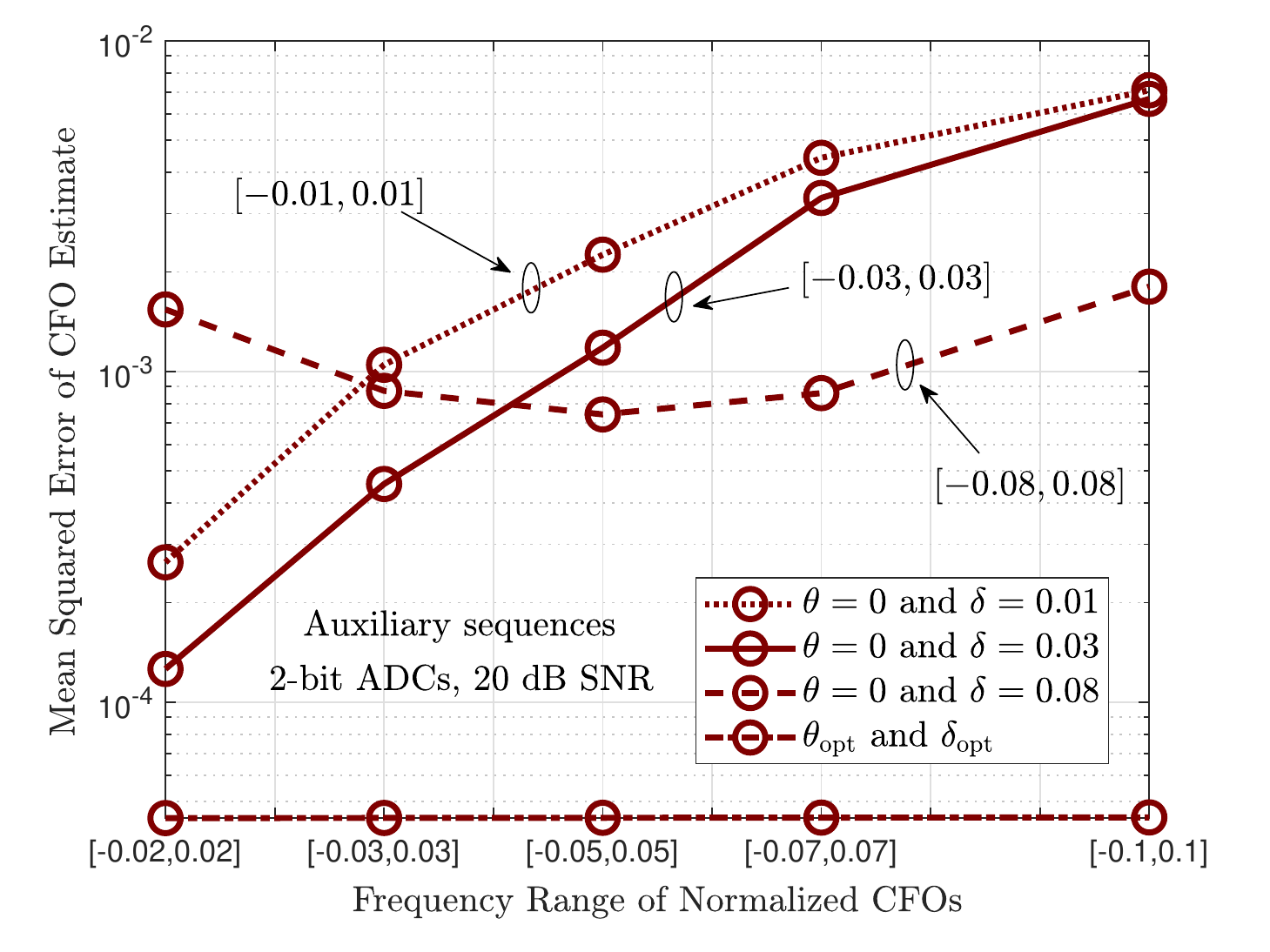}
\label{fig:simsubfigureMSECFOmutwobitb}}
\caption{Low-resolution frequency synchronization performance evaluations under various frequency ranges of normalized CFOs in a multi-user scenario. A total of $10$ UEs are assumed with each of them having a $2$-bit ADCs receiver. Their normalized CFOs are randomly distributed within the given frequency ranges. The SNR is $20$ dB. (a) Mean squared errors of the CFO estimates obtained via the ZC sequence based method and our proposed strategies against various frequency ranges of normalized CFOs. (b) Mean squared errors of the CFO estimates obtained via the proposed auxiliary sequences based method under various frequency ranges of estimation.}
\label{fig:simfigureMSECFOmutwobitabc}
\end{center}
\end{figure}

In this part of the simulation, we consider multiple UEs with $2$-bit ADCs equipped each. Other simulation parameters and the wideband channel model are the same as those in Section VI-B. In Fig.~\ref{fig:simfigureMSECFOmutwobitabc}, we evaluate our proposed methods assuming a total of $10$ UEs. Their normalized CFOs are randomly distributed within the given intervals $[-0.02,0.02]$, $[-0.03,0.03]$, $[-0.05,0.05]$, $[-0.07,0.07]$ and $[-0.1,0.1]$ (frequency ranges of normalized CFOs).

In Fig.~\ref{fig:simsubfigureMSECFOmutwobita}, we compare the CFO estimation MSEs between our proposed strategies and the ZC sequence based approach under various frequency ranges of CFOs. For the auxiliary and sum-difference sequences, we set the frequency range of estimation as $[-0.01,0.01]$ by configuring the double-sequence design parameters. As can be seen from Fig.~\ref{fig:simsubfigureMSECFOmutwobita}, our proposed methods can provide better CFO estimation performances for a wide range of frequencies in the multi-user setup.

In Fig.~\ref{fig:simsubfigureMSECFOmutwobitb}, we examine the impact of the frequency range of estimation on the frequency synchronization performances. In this plot, we only consider the auxiliary sequences based method. It is evident from Fig.~\ref{fig:simsubfigureMSECFOmutwobitb} that by increasing the frequency range of estimation from $[-0.01,0.01]$ to $[-0.03,0.03]$, the CFO estimation MSEs reduce. By further enlarging the frequency range of estimation to accommodate even more UEs (e.g., from $[-0.03,0.03]$ to $[-0.08,0.08]$), however, the overall CFO estimation performance degrades. These performance variations result from the fact that the frequency range of estimation ($\left[\theta-\delta,\theta+\delta\right]$ in this example) is not optimized according to the CFO distribution and the received SNRs. Finally, we plot the optimal case that uses $\theta_{\mathrm{opt}}$ and $\delta_{\mathrm{opt}}$ (obtained via (\ref{minmaxrft})) to construct the auxiliary sequences in Fig.~\ref{fig:simsubfigureMSECFOmutwobitb}. It can be observed that the corresponding CFO estimation performance is the best among all configurations and consistent across all frequency ranges of CFOs.
\section{Conclusions}
In this paper, we custom designed two double-sequence structures, i.e., auxiliary sequences and sum-difference sequences, for frequency synchronization in mmWave systems operating with low-resolution ADCs. The proposed two design options are different in terms of the ratio metric formulation, the double-sequence design parameters, and the achievable CFO estimation performance. We used analytical and numerical examples to show that our proposed methods are robust to the low-resolution quantization for a variety of network settings. We concluded from extensive empirical results that under low-resolution ADCs: (i) our proposed strategies provide promising overall frequency synchronization performance that can better trade off the CFO estimation accuracy and the frequency range of estimation, and (ii) our custom designed auxiliary and sum-difference sequences outperform the ZC sequences in estimating the frequency offset.
\begin{appendices}
\section{Proof of Lemma 1}\label{proof1}
For UE $u$, denote the signal power of the auxiliary channel $0$ output (i.e., $c^{0}_{u,\hat{b}}$) by $S^{0}_{u}$. According to (\ref{axchhh0}) and (\ref{axchhh1}), we can obtain
\begin{equation}\label{ax0outputsig}
S^{0}_{u}=\left[\mathbb{E}\left[\left|\left[\textbf{\textsf{H}}_{u}\left(e^{-\mathrm{j}(\theta\pm\delta)}\right)\right]_{\hat{b},:}\bm{f}\right|^{2}\right]\sin^{2}\left(\frac{N(\mu_u-\theta+\delta)}{2}\right)\frac{\sin^{2}\left(\frac{\mu_u-\theta-\delta}{2}\right)-\sin^{2}\left(\frac{\mu_u-\theta+\delta}{2}\right)}{\sin^{2}\left(\frac{\mu_u-\theta+\delta}{2}\right)\sin^{2}\left(\frac{\mu_u-\theta-\delta}{2}\right)}\right]^{2}.
\end{equation}
Denoting the quantization-plus-noise power of the auxiliary channel $1$ output (i.e., $c^{1}_{u,\hat{b}}$) for UE $u$ by $N^{1}_{u}$, we have
\begin{eqnarray}
N^{1}_{u}&=&N^2\left[\left(1-\kappa_{u}\right)^{2}\sigma_u^{2}+\sigma^{2}_{u,\mathcal{Q}}\right]^{2}\\
&=&N^{2}\left[\left(1-\kappa_{u}\right)\left(\kappa_{u}\mathbb{E}\left[\left|\left[\textbf{\textsf{H}}_u\left(e^{-\mathrm{j}(\theta\pm\delta)}\right)\right]_{\hat{b},:}\bm{f}\right|^{2}\right]+\sigma_u^2\right)\right]^{2}\label{ax1outputnoise}.
\end{eqnarray}
Assuming $\mu_u=\theta$, our proposed auxiliary sequences based estimator is unbiased \cite{phd}, i.e., $\mathbb{E}\left[\hat{\mu}_u\right]=\mu_u$ \cite{unbiasbook}. Note that the quantization distortion factor does not affect the unbiasedness of our estimator. The variance of the CFO estimate can then be expressed as \cite[equation~3.26]{phd}
\begin{equation}\label{ax01initvar}
\sigma^{2}_{\hat{\mu}_u,\mathrm{ax}}=\frac{1}{\varsigma^{2}_{\mathrm{ax}}S^{0}_{u}/N^{1}_{u}}\left[1+\mathcal{M}^{2}_{\mathrm{ax}}(\mu_{u})\right],
\end{equation}
where $\mathcal{M}_{\mathrm{ax}}(\mu_u)=\alpha_u$ for the proposed auxiliary sequences based method, and $\varsigma_{\mathrm{ax}}$ represents the slope of $\mathcal{M}_{\mathrm{ax}}(\cdot)$ at $\theta$, i.e.,
\begin{equation}\label{ax01moref}
\varsigma_{\mathrm{ax}} = \mathcal{M}'_{\mathrm{ax}}(\theta) = \frac{\sin(\delta)}{\cos(\delta)-1}.
\end{equation}
By plugging (\ref{ax0outputsig}), (\ref{ax1outputnoise}) and (\ref{ax01moref}) into (\ref{ax01initvar}), we can obtain (\ref{lllemma1}), which completes the proof.
\end{appendices}
\bibliographystyle{IEEEbib}
\bibliography{main_bib_WCL}
\end{document}